\documentclass{emulateapj}

\usepackage[OT1]{fontenc}
\usepackage{xspace} 

\usepackage[pdftex,pdfpagemode={UseOutlines},backref,breaklinks,bookmarks,bookmarksopen,colorlinks,linkcolor={black},citecolor={blue},urlcolor={red}]{hyperref}
\usepackage[all]{hypcap}

\usepackage{graphicx}
\usepackage{bm}
\usepackage{amsmath,amssymb}
\usepackage{aas_macros}
\usepackage{natbib}
\usepackage{bbold} 

\usepackage{multirow}

\let\oldtheequation\theequation
\makeatletter
\def\tagform@#1{\maketag@@@{\ignorespaces#1\unskip\@@italiccorr}}
\renewcommand{\theequation}{(\oldtheequation)}
\makeatother

\graphicspath{{figs-theory/}}

\def \spose#1{\hbox  to 0pt{#1\hss}}  
\newcommand {\lta} {\mathrel{\spose{\lower 3pt\hbox{$\sim$}}\raise  2.0pt\hbox{$<$}}}
\newcommand {\gta} {\mathrel{\spose{\lower  3pt\hbox{$\sim$}}\raise 2.0pt\hbox{$>$}}}

\newcommand {\kms} {\ifmmode  \,\rm km\,s^{-1} \else $\,\rm km\,s^{-1}  $ \fi }
\newcommand {\kpc} {\ifmmode  {\rm kpc}  \else ${\rm  kpc}$ \fi  }  
\newcommand {\pc} {\ifmmode  {\rm pc}  \else ${\rm pc}$ \fi  }  
\newcommand {\Msun} {\ifmmode {\rm M_{\odot}} \else ${\rm M_{\odot}}$ \fi} 
\newcommand {\Zsun} {\ifmmode {\rm Z_{\odot}} \else ${\rm Z_{\odot}}$ \fi} 
\newcommand {\yr} {\ifmmode yr^{-1} \else $yr^{-1}$ \fi} 
\newcommand {\hMsun} {\ifmmode h^{-1}\,\rm M_{\odot} \else $h^{-1}\,\rm M_{\odot}$ \fi}

\usepackage[usenames]{color}

\newcommand{\half}{\frac{1}{2}}

\newcommand{\ident}{\mathbb{1}} 
\newcommand{\kv}{\mathbf{k}}
\newcommand{\xv}{\mathbf{x}}

\newcommand{\pv}{\mathbf{p}}
\newcommand{\cv}{\mathbf{c}}
\newcommand{\thetav}{\bm \theta}
\newcommand{\muv}{{\bm \mu}}
\newcommand{\dirproc}{{\rm DP}}
\newcommand{\dpprec}{\mathcal{A}}

\newcommand{\psf}{{\Pi}}
\newcommand{\obscond}{\Omega}
\newcommand{\psfmodel}{p}

\newcommand{\ancdata}{\data_{\rm anc}}
\newcommand{\ancdatai}{\data_{{\rm anc},i}}
\newcommand{\condvar}{\mathcal{X}}
\newcommand{\ngal}{n_{\rm gal}}
\newcommand{\nepoch}{n_{\rm epoch}}
\newcommand{\fc}{\mathbf{H}}

\newcommand{\shear}{\gamma}
\newcommand{\conv}{\kappa} 
\newcommand{\normdist}{\mathcal{N}}
\newcommand{\shape}{e}
\newcommand{\summarystat}{\varepsilon}
\newcommand{\shapeint}{\shape^{\rm int}}
\newcommand{\shapeobs}{\summarystat^{\rm obs}}
\newcommand{\shapeobsmean}{\bar{\summarystat}^{\rm obs}}
\newcommand{\elensed}{\tilde{\shape}}
\newcommand{\galprops}{\omega}
\newcommand{\gravpot}{\Psi}
\newcommand{\gravpotinit}{\Psi^{\rm initial}}
\newcommand{\lenspot}{\psi}
\newcommand{\noiserms}{\sigma_{\rm pix}}
\newcommand{\noisermsni}{\sigma_{{\rm pix},n,i}}
\newcommand{\Sigmamat}{\mathsf{\Sigma}}

\newcommand{\numclasses}{M}

\def\sersic{S\'ersic}

\def\tractor{{\sc The Tractor}\xspace}
\def\emcee{{\sf emcee}\xspace}

\def\greatthree{{\sc GREAT3}\xspace}

\def\pr{{\rm Pr}}
\newcommand{\prf}[1]{\pr\left(#1\right)}
\def\data{{\mathbf{d}}}
\def\datamodel{\bar{\data}}

\def\kipac{Kavli Institute for Particle Astrophysics and Cosmology, Stanford University, 
Stanford, CA 94035, USA.
}
\def\slac{SLAC National Accelerator Laboratory, 
Menlo Park, CA 94025, USA.}
\def\ccin2p3{Centre de Calcul de l'IN2P3, USR 6402 du CNRS-IN2P3, 43 Bd. du 11 Novembre 1918, 69622 Villeurbanne Cedex, France.} 
\def\cmu{Department of Physics, Carnegie Mellon University, 
Pittsburgh, PA, 15213, USA.}
\def\llnl{Lawrence Livermore National Laboratory, 
Livermore, CA, 94551, USA.}
\def\davis{University of California, Davis, 
Davis, CA 95616, USA.}
\def\nyu{Center for Cosmology and Particle Physics, New York University, New York, NY, USA.}

\def\schneideremail{\tt schneider42@llnl.gov}

\begin{document}

\slugcomment{{LLNL-JRNL-661076}}
  
\title{Hierarchical probabilistic inference of cosmic shear}
\shorttitle{Hierarchical probabilistic inference of cosmic shear}
\shortauthors{M.~D.~Schneider et al.}

\author{Michael~D.~Schneider\altaffilmark{1, 2},
David~W.~Hogg\altaffilmark{3}, 
Philip~J.~Marshall\altaffilmark{4},\\
William~A.~Dawson\altaffilmark{1}, 
Joshua~Meyers\altaffilmark{5},
Deborah~J.~Bard\altaffilmark{4}, 
Dustin Lang\altaffilmark{6}
}

\altaffiltext{1}{\llnl}
\altaffiltext{2}{\davis}
\altaffiltext{3}{\nyu}
\altaffiltext{4}{\slac}
\altaffiltext{5}{\kipac}
\altaffiltext{6}{\cmu}

\email{\schneideremail}

\date{Draft \today}

\begin{abstract}
Point estimators for the shearing of galaxy images induced by gravitational lensing   
involve a complex inverse problem in the presence of noise, pixelization, and 
model uncertainties. 
We present a probabilistic forward modeling approach to gravitational lensing 
inference that has the potential to mitigate the biased inferences in most 
common point estimators and is practical for upcoming lensing surveys. 
The first part of our statistical framework requires specification of a 
likelihood function for the pixel data in an imaging survey given parameterized 
models for the galaxies in the images. 
We derive the lensing shear posterior by marginalizing over all intrinsic galaxy 
properties that contribute to the pixel data (i.e., not limited to galaxy ellipticities) 
and learn the distributions for the intrinsic galaxy properties via 
hierarchical inference with a suitably flexible conditional 
probabilitiy distribution specification.
We use importance sampling to separate the modeling of small imaging areas from the 
global shear inference, thereby rendering our algorithm computationally tractable for large surveys.
With simple numerical examples we demonstrate the improvements in accuracy from our importance 
sampling approach, as well as the significance of the 
conditional distribution specification for the intrinsic galaxy properties when the data are generated from an 
unknown number of distinct galaxy populations with different morphological characteristics.
\end{abstract}


\section{Introduction} 
\label{sec:introduction}

All observations to date are consistent with a cosmological model in which
mass is clustered on spatial scales ranging from galaxy sizes of order a 
kiloparsec to hundreds of megaparsecs, corresponding to angular scales on 
the sky of tens of arcseconds to several degrees. 
The cosmological mass distribution acts as a \emph{gravitational lens} for 
all luminous sources, which imparts inhomogeneous image distortions because 
the mass distribution is clustered. 
In principle, the coherent lensing distortions of luminous sources can be 
used to reconstruct the 3D cosmological mass distribution over most of the 
observable volume of the universe. 
However, except near the most dense and rare mass peaks, the gravitational lensing 
shearing of galaxy images is a percent-level effect perturbing unknown at order unity 
intrinsic galaxy shapes. The `cosmic shear' of galaxies therefore can 
only be detected through statistical correlations of large numbers of galaxy 
images.

Recently, cosmic shear has been used to place competitive constraints on
cosmological model parameters~\citep{jee2013,2013MNRAS.430.2200K,2013MNRAS.432.2433H,2014arXiv1408.4742M,2014MNRAS.442.1326K}. All analyses of which we are aware 
use point estimators for the 2D ellipticities of galaxies in a catalog to 
infer cosmic shear effects. From the estimated galaxy ellipticity components 
one can compute a two-point angular correlation function that is related to 
the mass clustering power spectrum in the standard cosmological framework.

There are several drawbacks to existing cosmic shear inference methods that 
may severely limit the amount of information that can be extracted 
about the 3D cosmological mass distribution and underlying cosmological model,
for a given set of observations~\citep{2012arXiv1211.0310L}. 
\citet{Bernstein+Armstrong2013} (BA14) give a review of 
the primary challenges in existing methods including low signal-to-noise, 
large instrumental and observational systematics, finite pixel sampling,
uncertainty in the morphologies of galaxies prior to lensing distortions, 
and biased ellipticity estimators in the presence of pixel noise.

As in BA14 and also \citet{sheldon2014}, \citet{Miller++2007,Miller++2013} we consider a probabilistic model for cosmic shear
inference that can in principle avoid many of the weaknesses in existing 
methods. 
Unlike in BA14 we initially set aside issues of 
computational feasibility and aim to specify a statistical 
framework that is a complete description of the data (i.e., including all 
physical parameters that are needed to completely describe the measured pixel 
values in a photometric survey). 
While we see value in simply writing down the complete statistical framework for cosmic shear
we were also able to identify statistical sampling and computational algorithms that are likely
to make our framework practical for upcoming surveys. We demonstrated a simple implementation 
of our framework in the \greatthree\footnote{\url{http://www.great3challenge.info/}, 
Our \greatthree submissions are under team `MBI': \url{http://great3.projects.phys.ucl.ac.uk/leaderboard/team/14}} 
community cosmic shear measurement challenge~\citep{Mandelbaum++2013} and further demonstrate the 
computational performance of our framework in subsequent papers~\citep{mbi-paper1,mbi-paper2}.

Most of the parameters that are needed to describe the pixel data are 
uninteresting for cosmology and can be marginalized out;
this is an advantage of probabilistic modeling.
If we can infer probability distribution functions (PDFs) 
for the nuisance parameters and marginalize---%
instead of either asserting distributions or else fixing the nuisance parameters
to heuristically chosen or maximum-likelihood values---%
we expect to obtain better (more accurate and also more conservative) inferences about the
cosmological parameters.
Adopting this probabilistic approach moves us (relative to other methods) along
the bias--variance trade-off towards less biased inferences.
Of course, in the context of fully probabilistic inference, without point estimators,
there isn't as clear a definition of ``bias'' and it may be difficult
to put the long literature on weak-lensing biases into context here.
However, we expect many of the known biases in point estimators to be ameliorated
when we permit the freedom to infer and marginalize out nuisances.
In detail, as we give more freedom to the model (more flexibility in the nuisance-parameter space),
we will move to even less biased (though also probably less precise) inferences,
possibly at large computational cost. 

Explicit specification or inference of the distributions of all model 
parameters not only has the potential to reduce biases in the shear 
inference but also is the only way to guarantee an optimal measurement 
(in the sense that no other accurate inference method could yield tighter marginal 
posterior distributions on the cosmological model parameters).
Most past cosmic shear analyses have not specified the distributions of 
galaxy intrinsic properties and observing conditions (e.g., the PSF) that are assumed in 
their cosmological inferences. 
But, because intrinsic galaxy properties and observing conditions do describe
important features of the data, all cosmic shear analyses must have 
(at least implicitly) marginalized over some distributions in these parameters.
Implicit marginalization over un-specified priors cannot yield an optimal 
measurement.
Said another way, without explicit marginalization of model parameters 
it is not possible to saturate the Cram\'{e}r-Rao bound~\citep[e.g.][]{kendallstuart}.
In particular, analyses in which measured ellipticities of galaxies are averaged together over sky patches (or something equivalent) have made an implicit (and strong) assumption about the distribution of intrinsic shapes; this assumption (since unexamined) is likely to 
be sub-optimal at best.

Once a complete statistical model is specified, there are three key practical 
challenges for probabilistic shear inference,
\begin{enumerate}
  \item Specification of an interim model describing galaxy morphologies, 
  flux profiles, and spectral energy distributions. 
  Galaxy features are complex and multi-faceted because of the evolution of galaxy properties with cosmic time, galaxy mergers, and varying environments.
  The \emph{observable} features of galaxies are further complicated by 
  object blending that may mimic intrinsic features of the galaxy population~\citep{dawson2014}.
  We require a parametric model for galaxies and a likelihood function 
  that can describe all such variations in galaxy images over the model parameter 
  ranges that are important for describing the data. 
  With an incomplete galaxy parametric model and without propagating all the 
  information about the model in the likelihood function, it is well known that 
  `model fitting biases' can be catastrophic for cosmic shear~\citep{Voigt+Bridle2010,2010A&A...510A..75M,Bernstein2010,2014MNRAS.441.2528K}. 
  \item Specification of the probability distribution for the galaxy model parameters to be marginalized in the shear inference.
  While the accurate specification of models for galaxy images may sound challenging 
  to the discerning reader, the specification of the joint distribution of 
  \emph{intrinsic} galaxy properties is perhaps even more daunting. 
  A primary aim of this paper is to describe and demonstrate how a hierarchical 
  model can allow meaningful inference of the properties of the galaxy distribution 
  simultaneously with the shear. 
  Without hierarchical inference (i.e., inferring the distributions of nuisance 
  parameters) we are susceptible to an `overconfidence problem' wherein asserting 
  a prior that is too simple can yield inferences that appear more precise than 
  warranted.
  \item Monte Carlo sampling of model parameters from the joint posterior distribution 
  given multi-epoch imaging of a large galaxy survey.
  Because both the shear and the PSF vary in correlated ways over any given image, 
  the inferences for different galaxies are statistically dependent. 
  In a straightforward probabilistic inference we would need to fit model parameters 
  simultaneously to all galaxy images, which is a formidable computational challenge. 
  Instead we sample model parameters for each galaxy independently and perform 
  a subsequent joint inference using importance sampling given interim posterior samples 
  for each galaxy image.
\end{enumerate}

In \autoref{sec:part1} we outline our framework for the complete cosmic shear inference problem.
In \autoref{sec:statistical_framework} we enumerate the components the framework and
in \autoref{sec:inference_methods} we outline the key algorithms we use for Monte Carlo sampling of
model parameters to render our framework computationally feasible.
In \autoref{sec:part2} we focus on implementation details for just the inference 
of intrinsic galaxy properties, leaving similar details for the PSF and lensing mass inferences 
to future work.
We describe our model for the distribution of intrinsic galaxy properties 
in \autoref{sub:dirichlet_process_priors_for_intrinsic_galaxy_properties}
and give numerical examples of shear and ellipticity inferences in \autoref{sec:numerical_example}.
In \autoref{sec:conclusions} we discuss how to expand the inferences demonstrated 
for the intrinsic ellipticity distribution to the PSF and lensing mass models and 
summarize our conclusions.
We refer the reader to~\citet{mbi-paper1} for further discussion of parameterized models 
for galaxy images and the integration of such models in our sampling framework.

\section{Part 1: Description of the statistical framework}
\label{sec:part1}

\subsection{Three conditionally independent branches}
\label{sec:statistical_framework}
We begin this section by enumerating the variables and dependencies of a complete 
statistical framework for shear inference. In subsequent sections we demonstrate the 
inference of a subset of the variables in our framework with numerical models. 
We defer a more detailed examination of some aspects of the statistical framework to later 
publications. But we believe it is useful at this stage to list all contributions to the 
cosmic shear inference problem given the considerable literature on the subject that has 
not yet converged on a unified statistical picture.

\begin{table}
\begin{center}
\caption{Sampling parameters for the full statistical model. 
The central line separates sampled from conditional parameters.}
\label{tab:sampling_parameters}
\begin{tabular}{cl}
\hline
Parameter & Description \\
\hline
$\thetav$ & Cosmological parameters \\
$\lenspot_{s}$   & 2D lens potential {\footnotesize (given source photo-$z$ bin $s$)} \\
$\psf_{i}$   & PSF in epoch $i$\\
$\obscond_{i}$ & Observing conditions in epoch $i$\\
$\{\galprops_n\}$ & Galaxy model parameters; $n=1,\dots,\ngal$\\
$\{\alpha_n\}$ & Parameters for the distribution of $\{\galprops_n\}$\\
$\{\xi_n\}$    & Scaling parameters for $\{\galprops_n\}$\\
$m,\tau$ & Hyperprior parameters for $\{\xi_n\}$ \\
$\dpprec$       & Hyperparameter for $\{\alpha_n\}$ classifications\\

\hline
$\{\data_n\}$ & Pixel data for galaxies $n=1,\dots,\ngal$ \\
$G_0|a_{\eta}$     & Prior specification for $\{\alpha_n\}$\\ 
$s$       & Source sample (e.g., photo-$z$ bin) \\
$W$       & Survey window function \\
$\ancdatai$ & Ancillary data for PSF in epoch $i$\\
$\psfmodel$  & Prior params. for observing conditions\\
$a$       & Prior params. for $\dpprec$ \\
$\noiserms$ & Pixel noise r.m.s.\\
$I$       & Model selection assumptions \\
\hline
\end{tabular}
\end{center}
\end{table}
\begin{figure}[htb]
  \centerline{
    \includegraphics[width=0.45\textwidth]{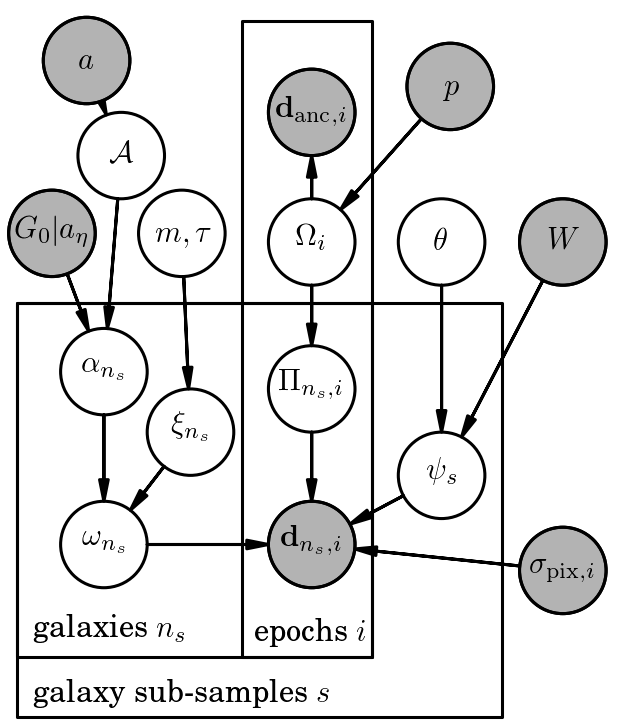}
  }
  \caption{Probabilistic graphical model for our complete framework for shear inference.
  Arrows indicate statistical dependencies. Grey shading indicates quantities 
  that are not sampled. 
  For the lens potential $\lenspot$ we include only the dependencies in the 
  marginal distributions after integrating out the 
  line-of-sight distribution uncertainties contributing to the lens 
  potential.
  }
  \label{fig:model_pgm}
\end{figure}

\subsubsection{Lensing mass distribution}
We start by specifying the parameters $\thetav$ of the cosmological model,
sampled from a prior distribution $\pr(\thetav)$ given all past cosmological experiments.
Although not explicitly implemented in this paper, we expect $\thetav$ to include 
parameters such as the mean mass density $\Omega_{m}$ and the r.m.s. of 
mass density fluctuations $\sigma_{8}$ that primarily determine the amplitude of the 
cosmic shear correlation function.

From these parameters, we can predict the 3D cosmological mass distribution, described 
by the 3D gravitational lensing potential $\gravpot$. 
Assuming, for example, Gaussian initial conditions $\gravpotinit$ for the 3D mass density 
in the early universe and gravitational evolution according to General Relativity,
the probability distribution for the late-time 3D mass density $\gravpot$ responsible 
for gravitational lensing of galaxies depends 
only on the cosmological parameters and the initial conditions, 
$\pr(\gravpot|\thetav, \gravpotinit)$.
We discuss possibilities for inferring constraints on $\gravpot$ in 
\autoref{sub:a_consistent_shear_model_for_the_whole_sky} but otherwise 
confine our analyses to the inference of gravitational lensing quantities 
after projecting the 3D gravitational potential $\gravpot$ over the line-of-sight distribution 
of lenses in a survey.

We define the 2D lensing potential that describes the shearing (and magnification) of any 
sample of galaxies in a survey by the following integral of the 
3D potential $\gravpot$~\citep[see, e.g., equation 6.14 of][]{bartelmann01}
\citep[also section 3.2 of][]{1996astro.ph..6001N},
\begin{equation}
  \bar{\lenspot}_{s}(\xv) \equiv
  W(\xv)
  \int da\,
  \gravpot(\xv,a)
  K(a ; A_s),
\end{equation}
where $a$ is the cosmological scale factor in the Freidmann equation, 
$\xv$ are the coordinates in the plane of the sky,
$W(\xv)$ is the angular window function of the survey,
the subscript $s$ denotes a particular sample of `source' galaxies 
that are lensed by the foreground potential $\gravpot$, 
and
$K$ is the lensing kernel including the lensing efficiency based on the 
distances between observer, lens, and sources, and the integral over the 
source distribution with parameters $A_s$~\citep[see equation 6.19 of][]{bartelmann01}.

The parameters $A_s$ can include the uncertainties in the 
survey redshift distribution \citep[e.g., from photometric redshift errors, see][]{1986ApJ...303..154L,1995AJ....110.2655C,2006MNRAS.366..101H} as well as 
other astrophysical systematics that are redshift-dependent 
\citep[such as intrinsic alignments of galaxies, see][]{2001MNRAS.320L...7C,bridleking07}. 
Marginalizing the nuisance parameters in the 
line-of-sight projection, the 2D lensing potential $\lenspot$ given the cosmology and 
survey sample is 
distributed according to,
\begin{multline}\label{eq:lenspotential}
  \pr(\lenspot_s | \theta, s, W) \propto
  \int d\gravpot\,
  \pr(\gravpot | \thetav) 
  \\
  \times \int dA_s\, 
  \pr(\lenspot_s | \gravpot, A_s, W)
  \pr(A_s | \thetav, s, W).
\end{multline}
For a deterministic cosmological model~\citep[as emphasized in][]{2013MNRAS.432..894J}, 
$\pr(\lenspot | \gravpot, A, W) \propto
\delta_{\rm D} \left(\lenspot - \bar{\lenspot}(\gravpot, A, W)\right)$, where 
$\delta_{\rm D}$ is the Dirac delta function.

\subsubsection{Point-spread functions}
To connect the lens potential to the observable properties of galaxy images, 
we  must also specify the distribution of the point-spread function (PSF)
realizations specific to the (time-variable) 
observing conditions at epoch $i$ given parameters $\obscond_i$ such as 
the seeing, optics alignments, and detector response,
$\pr(\psf_{i} | \obscond_{i})$.
For given $\obscond$, the PSF $\psf(\fc_n; \obscond)$ 
can be computed at the focal plane position $\fc_n$ 
for any galaxy $n$.
The observing conditions $\obscond_i$ in an observation epoch $i$ 
will not be known perfectly but will be constrained by ancillary 
data $\ancdata$ and a distribution $\pr(\obscond_i | \psfmodel)$ conditioned 
on time-independent PSF model parameters $\psfmodel$ (e.g., the 
allowed distributions of optics distortions or atmospheric turbulence 
power spectra). 
The complete model for the PSF at all field positions $\fc_n$ 
in observation epoch $i$ is not separable for different field positions 
after marginalizing over the observing conditions,
\begin{multline}\label{eq:prob_psf_model}
  \pr(\left\{\psf_{n,i}\right\}_{n=1}^{\ngal} | 
  \left\{\fc_{n}\right\}_{n=1}^{\ngal}, \ancdatai, \psfmodel) =
  \\
  \int d\obscond_i\,
  \left[
  \prod_{n=1}^{\ngal}
  \pr(\psf_{n,i} | \fc_{n}, \obscond_i)
  \right]
  \\
  \times
  \pr(\ancdatai | \obscond_i) 
  \pr(\obscond_i | \psfmodel).
\end{multline}
where $\pr(\ancdatai | \obscond_i)$ is the likelihood of the 
ancillary data given the specific observing conditions in epoch $i$.
We will refer to \autoref{eq:prob_psf_model} as our `probabilistic PSF model'.

Note however, that we infer a distinct PSF realization $\psf_{n,i}$ for each 
galaxy $n$ in each observation epoch $i$, which are related across 
galaxies in a given epoch by the common observing conditions $\obscond_{i}$.
The PSFs in different epochs are further related by another level of hierarchy 
with a common PSF model with parameters $\psfmodel$.
So, we can also directly marginalize the likelihood over the PSF realizations,
\begin{align}
  &\pr(\data | \{\galprops_n\}, \lenspot, \psfmodel) = 
  \prod_{n=1}^{\ngal}
  \prod_{i=1}^{\nepoch}
  \int d\obscond_{i}\, \pr(\obscond_{i} | \psfmodel)
  \notag\\
  &\times
  \int d\psf_{n,i}\, \prf{\psf_{n,i} | \obscond_{i}}
  \prf{\data_{n,i} | \galprops_n, \lenspot, \psf_{n,i}}
  \notag\\
  &= 
  \prod_{i=1}^{\nepoch}
  \int d\obscond_{i}\, \pr(\obscond_{i} | \psfmodel)
  \left[
  \prod_{n=1}^{\ngal}
  \prf{\data_{n,i} | \galprops_n, \lenspot, \obscond_{i}}
  \right]
  \label{eq:psf_marg_like}\\
  &= \prod_{i=1}^{\nepoch}
  \prf{\data_{i} | \{\galprops_n\}, \lenspot, \psfmodel}
  \\
  &\neq
  \prod_{n=1}^{\ngal}
  \prf{\data_{n} | \galprops_n, \lenspot, \psfmodel}
\end{align} 
Marginalizing the PSF realizations $\psf_{n,i}$ over all galaxies $n$ 
in a given epoch $i$ still yields a likelihood function that is separable 
for separate objects $n$~\autoref{eq:psf_marg_like}, conditioned on the 
observing conditions $\obscond_{i}$.
This shows that more information about the observing conditions $\obscond_{i}$
leads to smaller statistical correlations among the marginal likelihoods for 
different galaxies independent from the actual PSF realizations. Under our 
framework therefore, it is the parameters of the distribution from which PSF 
realizations are drawn that are most important to constrain rather than estimation 
and interpolation of PSF realizations at different image plane locations.

\subsubsection{Intrinsic source properties}
Finally, the intrinsic properties $\galprops_n$ of a model for galaxy $n$ (e.g., ellipticity, size, brightness) 
are described by the distribution 
\begin{equation}
  \pr(\galprops_{n} | \alpha, I),
\end{equation}
where $\alpha$ are parameters in the distribution of the intrinsic galaxy properties and $I$ denotes 
galaxy modeling assumptions (such as the form of the galaxy morphology parameterization). 
For example, $\alpha$ could describe the width of the intrinsic (i.e., unlensed) ellipticity distribution 
of a given galaxy population. And in~\citet{mbi-paper1}, the model assumptions $I$ include modeling all galaxies 
as elliptical \sersic~profiles.
It will be 
important to keep $I$ explicit as we consider possible model-fitting biases in 
our framework~\citep{Voigt+Bridle2010,2014MNRAS.441.2528K,mbi-paper1}.

\subsection{Sampling methods to divide and conquer}
\label{sec:inference_methods}

\subsubsection{Interim sampling of model parameters for individual sources} 
\label{sub:likelihood_specification}

The pixel data $\data_n$ in the vicinity of a galaxy $n$ enters our framework only through the likelihood function 
\begin{equation}\label{eq:data_cond_dist}
  \pr(\data_n | \galprops_n, \lenspot, \psf).
\end{equation}
We can reasonably assume uncorrelated noise in the common case that sky noise (i.e., Poisson noise from sky background) dominates the noise in the pixel values, 
giving a likelihood function,
\begin{equation}\label{eq:likelihood}
  \ln \pr(\data_n) =
  -\half
  \sum_{i=1}^{n_{\rm epochs}}
  \frac{\left(\data_{n,i} - \datamodel\left(\galprops_n, \lenspot, \psf_{i}\right) \right)^2}{\noisermsni^2} 
  + {\rm const.},
\end{equation}
where $\datamodel$ is the model prediction for the pixel values,
$i$ indexes multiple observation epochs,
and $\noisermsni$ is the noise r.m.s. per pixel for object $n$ in epoch $i$.
Note the model prediction in the likelihood includes the PSF $\psf_i$ for 
each observation $i$ of a given galaxy $n$. This makes the model predictions 
potentially expensive to compute unless the convolution of the PSF with the 
galaxy model can be done efficiently. We will return to this issue in 
\autoref{sub:modeling_galaxy_images_the_tractor} and \citet{mbi-paper1}. 
We discuss the alternative proposed by BA14 of 
using moments of the galaxy intensity profiles in place 
of pixel values in Appendix~\ref{sec:galaxy_moments}.

In \autoref{eq:data_cond_dist} we have assumed that the likelihood function for all sources 
in an image can be factored into a product of likelihoods for distinct sources. This will 
not be true in when the pixel noise is correlated for sources that are adjacent on the sky 
(e.g., when unresolved flux contaminates the pixels of neighboring sources).

The model prediction $\datamodel$ for the pixel values of a galaxy image depends on the cosmological mass density $\lenspot$ through the action of 
gravitational lensing on the galaxy morphology and flux (in any given bandpass as lensing is achromatic). 
For simple elliptical models of galaxies, the ellipticity parameter 
in $\galprops_n$ is degenerate with the reduced shear 
$g \equiv \shear / (1 - \conv)$. However, the statistical distribution of 
all elements of $\galprops$ should be isotropic across different galaxies in 
the sky (although this is not true for small fields, e.g., around a cluster), while the ellipticities of lensed galaxy images are not.
We therefore maintain a conceptual distinction between 
$\galprops_n$ and $\lenspot$ in the model for pixel 
values of a galaxy image $\datamodel$ even though this distinction 
is not computationally meaningful for the simplest models of galaxy 
images. See also the discussion following  \autoref{eq:importance_samples} below.

The first step in our computational algorithm is to draw 
`interim samples' of $(\galprops_n, \lenspot, \psf_{n,i})$ 
via Markov Chain Monte Carlo (MCMC) from the 
likelihood in \autoref{eq:likelihood} for a single source (or region of sky)
identified in all available exposures of a survey. The method of source 
identification for selecting the pixel values $\data_n$ in 
\autoref{eq:likelihood} is not important as long as our model 
predictions for the observed pixels allow for observationally truncated source 
profiles or multiple overlapping sources (as will be needed for blended 
objects). Allowing for such selection effects may also admit interim 
sampling of the combined pixel data from different instruments or surveys.

\paragraph{Lensing mass inference} 
\label{sub:a_consistent_shear_model_for_the_whole_sky}

The intrinsic galaxy properties $\galprops_n$ and the PSF $\psf_{n,i}$ can be 
interim sampled with distinct parameters for each object. 
We might consider interim sampling uncorrelated shears and magnifications for every object,
but in most cases the shear and the intrinsic ellipticity are strongly degenerate for 
isolated objects (the exception being resolved galaxy images with multiple morphological components
that have different responses to the action of an applied shear). 
To infer a shear or lens potential that varies over the sky, 
we then need to specify the model correlations between shears at different sky locations and 
redshifts. This is distinct from conventional algorithms that involve cross-correlating 
galaxy ellipticity estimators at different sky locations. We instead seek a generative 
model for the shear correlations. 

We can infer variable shear models without a cosmological prediction of the 
mass density (which can be computationally expensive) by specifying an interim 
prior for the lens potential $\lenspot_s$ for a given source distribution.
The only requirement is that the mathematical specification of the 
interim prior for $\lenspot_s$ allows for spatial correlations that encapsulate the possible 
cosmological interpretations. 

In practice, we first sample galaxy model parameters without a correction for the 
applied shear and magnification. Given the galaxy model parameters, we can then 
reinterpret the parameters under an assumed lens potential model because we 
know how lensing affects the model for the galaxy light profile. 
We therefore ignore the lensing parameters in the first step of our inference,
including a model for lensing shears and magnifications only when combining 
inferences from distinct sources as we describe next.

\subsubsection{Hierarchical inference via importance sampling} 
\label{sub:importance_sampling}
It is possible to infer model constraints independently for 
each galaxy and then combine these independent inferences in a hierarchical 
framework using the technique of 
`importance sampling'~\citep[e.g.][]{geweke1989,wraith++2009,2010ApJ...725.2166H}.

Our goal with the importance sampling 
algorithm is to evaluate the likelihood marginalized over individual 
galaxy intrinsic properties and PSF realizations\footnote{We may not always want to 
assume that the PSF realizations for a given galaxy are statistically independent across 
observation epochs, but it is a useful clarifying assumption here.},
\begin{multline}\label{eq:marg_likelihood}
  \pr(\data | \alpha, \obscond, \lenspot)
  \propto
  \prod_{n=1}^{\ngal}
  \int d\galprops_{n}\,\pr(\galprops_n | \alpha)
  \\
  \times 
  \prod_{i=1}^{\nepoch}
  \int   d\psf_{n,i}\,\pr(\psf_{n,i} | \obscond_{i})
  \\
  \times
  \pr(\data_{n,i} | \galprops_{n}, \lenspot, \psf_{n,i})
  ,
\end{multline}
where 
\begin{equation}
  \data \equiv \left\{\left\{\data_{n,i}\right\}_{i=1}^{\nepoch}\right\}_{n=1}^{\ngal},
\end{equation}
and $\obscond\equiv \left\{\obscond\right\}_{i=1}^{\nepoch}$.
Using the identity,
\begin{equation}
  \int dx\, p(x) f(x) = \int dx\, p(x) g(x) \frac{f(x)}{g(x)},
\end{equation}
for arbitrary probability distributions $p(x), f(x)$ and
assuming $g(x)$ has nonzero probability mass over the domain where $f(x)$ is nonzero,
we can factor \autoref{eq:marg_likelihood} as,
\begin{multline}\label{eq:marg_like_IS_identity}
  \pr(\data | \alpha, \obscond, \lenspot)
  \propto
  \prod_{n=1}^{\ngal}\prod_{i=1}^{\nepoch}
  \int d\galprops_n\int d\psf_{n,i}\,
  \\
  \times
  \frac{\pr(\galprops_n|\alpha)}{\pr(\galprops_n|I_0)}
  \frac{\pr(\psf_{n,i}|\obscond_{i})}{\pr(\psf_{n,i}|I_0)}
  \\
  \times
  \pr(\data_{n,i} | \galprops_n, \lenspot, \psf_{n,i}) 
  \pr(\galprops_n | I_0)
  \pr(\psf_{n,i} | I_0).
\end{multline}
The final line of \autoref{eq:marg_like_IS_identity} 
defines the `interim posterior' from which we draw samples when 
we analyze each galaxy independently~\citep[following][]{2010ApJ...725.2166H}. 
We refer to 
$\pr(\galprops_n | I_0)$ and $\pr(\psf_i | I_0)$ as `interim priors'. 
The interim priors can be chosen to make computations convenient. 
We have found flat or broad Gaussian distributions to be sufficient interim 
prior specifications.

We use the following algorithm to combine samples from the interim posteriors 
for each galaxy into a hierarchical inference of the global marginal posterior.
For each object $n$ we generate $K$ samples $\left[\galprops_{nk}, \left\{\psf_{nik}\right\}_{i=1}^{\nepoch}\right]$, 
sampled from the interim posterior given pixel data for galaxy $n$ in all epochs,
\begin{multline}\label{eq:interim_posterior}
  \prf{\galprops_n, \psf_{n} | \data_{n}} = \frac{1}{Z_{n}} 
  \\
  \times
  \prf{\data_{n}|\galprops_n, \psf_{n}} \prf{\galprops_n | I_0}
  \prf{\psf_{n} | I_0},
\end{multline}
where $Z_{n}$ is a normalization that we never need to compute, 
and $I_0$ denotes the set of assumptions encoded into the interim priors.
Once we have this $K$-element interim sampling for every galaxy $n$ 
we can build importance-sampling approximations to various other 
marginalized likelihoods. 
Specifically, for the integral in \autoref{eq:marg_like_IS_identity}, 
the marginalized likelihood can be approximated by 
\begin{align}
  \pr(\data_{n}|\alpha, \obscond, \lenspot) &\approx \frac{Z_{n}}{K}
  \sum_{k} 
  \frac{\pr(\galprops_{nk} | \alpha, \lenspot)}{\pr(\galprops_{nk} | I_0)}
  \frac{\pr(\psf_{nk} | \obscond)}{\pr(\psf_{nk} | I_0)},
  \label{eq:importance_samples}
  \\
  \pr(\data | \alpha, \obscond, \lenspot) &= 
  \prod_{n=1}^{\ngal} 
  \pr(\data_{n} | \alpha, \obscond, \lenspot).
  \label{eq:marg_like}
\end{align}
What we have achieved with \autoref{eq:importance_samples} is the ability to fit 
galaxy models and PSFs to individual galaxies (via MCMC) and then to combine the 
fit information from each galaxy into inferences on the distributions of intrinsic 
galaxy properties $\alpha$ and PSF parameters $\psfmodel$. \autoref{eq:importance_samples} is 
fast to compute once the $K$ samples for each galaxy are generated.
For the final shear inference we will marginalize the parameters $\alpha$, but 
with \autoref{eq:importance_samples}
we have already addressed the primary computational bottleneck in modeling images of 
large numbers of sources in a field.

Although not part of the preceding derivation, we inserted the 
lensing potential $\lenspot$ into the list of conditional dependencies in 
the numerator of \autoref{eq:importance_samples}. This is because given interim 
posterior samples of galaxy model parameters, we can always reinterpret those 
model parameters under a different assumption for the shear for any model where we 
know how to calculate the action of shear on the model galaxy\footnote{This is a distinct concept 
from estimating the action of shear on the observed pixel values of a galaxy as in BA14.}.

To evaluate \autoref{eq:importance_samples} we need $K$ independent samples 
from the interim posteriors for each galaxy. While this can be slow to generate via 
MCMC when $K$ is required to be large, we have found in simple tests that even 
$K=1$ may be sufficient when assuming a constant shear over a given area of the 
sky~\citep{mbi-paper2}. 
We use $K=10$ in the numerical examples in \autoref{sec:part2}.


\paragraph{Importance sampling the lens potential}

Given interim samples of the lens potential $\lenspot_s$
we can then infer cosmological parameter constraints from the sampled lens potentials for 
all galaxy samples in all fields using importance 
sampling as in \autoref{eq:importance_samples}, but with a new conditional PDF
that includes the dependence of the lens potential cosmological parameters $\thetav$ and 
the cosmological initial conditions $\gravpotinit$,
\begin{equation}\label{eq:lens_pot_importance_sampling}
  \pr(\data_n | \gravpotinit, \theta, W) \propto 
  \frac{1}{N}
  \sum_{k=1}^{N} 
  \frac{\prod_{s} \pr(\hat{\lenspot}_{s,k} | \gravpotinit, \theta, W)}
  {\prod_{s}\pr(\hat{\lenspot}_{s,k} | I)},
\end{equation}
where, again, we have implicitly marginalized over the line-of-sight distribution error parameters 
$A_{s}$ from \autoref{eq:lenspotential}.

\autoref{eq:lens_pot_importance_sampling} also helps to understand the utility in 
separating the galaxy population into different samples $s$. It is straightforward 
to derive separate inferences of $\gravpotinit$ and $\thetav$ for different 
sub-samples $s$ of the galaxies to test for consistency with respect to unmodeled 
systematic errors or new physical phenomena not captured by the cosmological model 
for the mass density evolution. 
One could similarly infer different distributions of the intrinsic galaxy properties $\galprops$
for different sub-samples to test for variable model fitting biases and 
unmodeled redshift evolution in the galaxy population. But, once we have computed the inferences 
of the 2D lens potential for different sub-samples $s$, the combined inference 
of cosmology in \autoref{eq:lens_pot_importance_sampling} is not much more complicated 
than that for an undivided galaxy sample. 

An algorithm as in \autoref{eq:lens_pot_importance_sampling} could also obviate 
the need to compute covariances 
of cosmological correlation functions between tomographic redshift bins, which is estimated 
to be a formidable challenge for upcoming surveys~\citep{2013PhRvD..88f3537D,2013MNRAS.432.1928T,2013JCAP...11..009M}

To summarize, by placing an interim prior on $\lenspot$ that includes spatial correlations over 
the observed sky area, we stand to gain from 
a similar computational factorization of the analysis as that achieved with the 
model for the intrinsic galaxy properties. 
Most published cosmic shear 
analyses follow a reverse algorithm where the shear is estimated, turned into 
two-point correlation estimators and compared with a two-point correlation model. 
Instead, our framework must start with the two-point correlation model (for a Gaussian 
density field), realize the shear field, apply the realized shear to each 
galaxy model and then compare the pixel model with the data for each galaxy.
We consider the computational separation of each of these steps to be a 
key benefit of the framework we present here.

Finally, one should keep in mind that the relationship between the lens potential 
defined over the sky and the lensing distortions of measured galaxy images are 
dependent on the astrometric solution for each exposure. However, we expect this to be a 
negligible contribution to the shear inference in most cases.
Possible centroid shifts of galaxy images should instead be captured by the 
PSF model.

\subsection{Statistical framework summary} 
\label{sec:full_model}

In Sections~\ref{sec:statistical_framework} and \ref{sec:inference_methods} we 
introduced a number of parameters with nontrivial interdependencies. 
Here we collect all the parameters we have introduced and 
summarize the statistical framework by presenting the 
factorization of the unmarginalized joint posterior for all sampled model parameters,
\begin{multline}\label{eq:full_model}
  \pr(\thetav, \{\lenspot_s\}, \{\psf_{i}, \obscond_{i}\}, \{\galprops_n, \alpha_n, \xi_n\}, \dpprec,
  \tau, m, a_{\eta} |\{\data_n\}, \condvar 
    ) \propto\\
  \prf{\thetav} \cdot \prf{\{\lenspot_{s}\} | W, \{s\}, \thetav}
  \cdot \pr(\dpprec | a)
  \\
  \times
  \prod_{n=1}^{\ngal} 
  \pr(\galprops_n | \alpha_n, I) 
  \cdot
  \pr(\alpha_n | \dpprec, a_{\eta}) \cdot
  \pr(\xi_n | m, \tau)\\
  \times
  \prod_{i=1}^{\nepoch}
  \prf{\psf_{n,i} | \obscond_{i}, I) \cdot \pr(\obscond_{i} | \psfmodel, \ancdatai}
  \\ \times
  \prf{\data_{n,i}|\galprops_n, \xi_n, \lenspot_{s}, \psf_{n,i}}
\end{multline}
where in the first line we collapsed the conditional parameters into,
\begin{equation}
  \condvar\equiv \left[\ancdata, \{s\}, W, a, \psfmodel, I\right].
\end{equation}
We collect the definitions of all variables in \autoref{eq:full_model}
in \autoref{tab:sampling_parameters} and display the statistical dependencies 
in \autoref{fig:model_pgm}.

The final line of \autoref{eq:full_model} is the likelihood function 
for the pixel data $\data_{n,i}$ of galaxy $n$ in observation epoch $i$, which 
depends on the intrinsic galaxy properties $\galprops_n, \xi_n$, the 
lensing potential $\lenspot_{s}$ for source sample $s$ containing galaxy $n$, 
and the PSF $\psf_{n,i}$. The preceding lines in \autoref{eq:full_model} 
describe the hierarchical PDFs for the conditional dependencies in the 
likelihood function, including the important factorizations across 
distinct galaxies and observation epochs.
In \autoref{sec:discussion} we discuss how marginalization of the 
conditional dependencies in the likelihood couples the inference 
from individual galaxies and epochs as determined by the hierarchical 
parameters that are constant across galaxies and epochs in \autoref{eq:full_model} 
(e.g., parameters that do not have $n$ or $i$ subscripts).

Note also we assumed in \autoref{eq:full_model} that the errors in the 
line-of-sight distribution (e.g., photo-$z$ errors) 
are already marginalized as in \autoref{eq:lenspotential}. We reserve 
a more thorough exploration of line-of-sight parameter modeling for future work.


\section{Part 2: Implementation for inferring intrinsic galaxy properties}
\label{sec:part2}

\subsection{Model for the conditional distribution of intrinsic galaxy properties} 
\label{sub:dirichlet_process_priors_for_intrinsic_galaxy_properties}

We choose a non-parametric distribution to describe the galaxy 
model parameters
both because we have little information to guide us on the choice of a 
parametric distribution and because we need a flexible distribution to minimize 
any biases in the model parameter inferences (as mentioned in \autoref{sec:introduction}).
We therefore choose a Dirichlet Process (DP)
model~\citep[e.g.][]{ferguson1973,RSSB:RSSB190,neal00,muller2004,wang2011} 
for the distribution of intrinsic galaxy properties, with 
distinct parameters $\alpha_n$ for each galaxy $n$. 

Assuming a Normal (i.e., Gaussian) distribution for the 
intrinsic galaxy properties $\galprops$,
the DP model can be represented by the following 
hierarchy~\citep[][Eqn. 2.1]{neal00},
\begin{equation}\label{eq:dpdef}
  \galprops_n \sim \normdist(0, \alpha_n),\quad
  \alpha_n \sim G(\alpha_n | \dpprec),
  \quad
  G \sim \dirproc(\dpprec, G_{0}),
\end{equation}
where $\normdist$ indicates the Normal distribution with mean and (co)variance parameters.
We assume a Normal distribution for $\galprops_n$ in \autoref{eq:dpdef} both 
for specificity and because it is sufficient for all our numerical studies in 
\autoref{sec:part2}.
\autoref{eq:dpdef} indicates that for every galaxy $n$ the properties $\galprops_n$ of that galaxy
(such as ellipticity, size, and flux) are Gaussian distributed with 
variance parameters $\alpha_n$. The subscript on $\alpha_n$ indicates we allow for 
\emph{different} variances for every galaxy. Without a statistical model that 
relates inferences of $\alpha_n$ for different $n$ we would then be assuming that 
every galaxy in the universe is realized from a distinct generative (i.e., physical) 
mechanism from every other galaxy. This is of course not consistent with our 
understanding of cosmology and galaxy formation. Instead we introduce 
relationships between $\alpha_n$ parameters for different galaxies $n$ by means 
of the DP model.

Another way to think about \autoref{eq:dpdef} is that we 
specify a distribution on $\galprops$ that is a mixture of distributions 
with parameters $\alpha\sim G_0$
with a hyperprior on the mixing proportions given by the Dirichlet 
distribution.

For sampling algorithms 
it is useful to first marginalize the 
distribution $G$ in \autoref{eq:dpdef} to get the conditional 
updates~\citep[eq. 2.2 of][]{neal00},
\begin{multline}\label{eq:dp_cond_prior}
  \alpha_n | \alpha_1,\dots,\alpha_{n-1} \sim
  \frac{1}{n-1+\dpprec} \sum_{\ell=1}^{n-1} \delta_{\rm D} \left(\alpha_{\ell}\right)
  \\
  +
  \frac{\dpprec}{n-1+\dpprec}G_{0}(\cdot),
\end{multline}
where $n$ indexes galaxies and $\delta_{\rm D}$ is the Dirac delta function.
Given the discreteness property of the DP there is nonzero probability that draws of $\alpha_n$ will be repeated.
Let $\alpha^{*}_1,\dots,\alpha^{*}_{K}$ be the unique values among 
$\alpha_1,\dots,\alpha_{n-1}$, and let $N_c$ be the number of repeats of $\alpha^{*}_c$ in the ``latent class'' $c$.
Then the conditional update distribution can be equivalently written as
\citep[eq.~10 of][]{Teh2010}
\begin{multline}\label{eq:dp_cond_prior_v2}
	\alpha_n | \alpha_1,\dots,\alpha_{n-1} \sim \frac{1}{n-1+\dpprec} \\
	\left(\sum_{c=1}^{K}N_c \delta_{\rm D}\left(\alpha^{*}_{c}\right)
	+ \dpprec G_{0}\left(\cdot\right) \right).
\end{multline}
This equation is useful because it shows the clustering properties of the DP
and the meaning of the parameter $\dpprec$.
The value of $\alpha^{*}_c$ will be assigned to $\alpha_{n}$ with probability
proportional to the number of times $\alpha^{*}_c$ has be previously drawn, $N_c$.
Note that when $\dpprec$ is large, successive values for $\alpha_n$ are 
drawn from the base distribution $G_0$ with high probability, generating a new 
$\alpha^{*}_{K}$ each time. 
When $\dpprec$ is small, $\alpha_n$ is assigned to one of the previous 
values of $\alpha^{*}_c$ with high probability. 
We will refer to groups of galaxies $\{n\}$ that have equal values of $\alpha^{*}_k$
as a `cluster' or `sub-population', representing potential galaxy 
type classifications 
(analogous to the common `early-' or `late-type' classifications).
\autoref{eq:dp_cond_prior} shows that larger values of $\dpprec$ will lead 
to more clusters or categories of galaxy types while bringing $\dpprec$ to 
zero will force all galaxies to be assigned to the same cluster or type classification.
Note that as long 
as the observations are exchangeable (which is true for sources in an image), 
the ordering of values in $\alpha$ is arbitrary.

To summarize, we are motivated to use the DP model for the following reasons,
\begin{itemize}
  \item The DP is analogous to 
  a mixture model when the number of mixture components is learned from the data, 
  \item Samples for galaxy $n$ are naturally informed by those for galaxies $1,\dots, n-1$, 
    thereby improving the sampling efficiency (for arbitrary ordering of observed galaxies),
  \item The DP can find unknown clusters in the galaxy parameters, 
  teaching us about galaxy formation and different prospects for shear inference from different data subsets. For example, we show in \autoref{sub:dp_inference} that knowledge of a
  galaxy sub-population that is more round than average can improve the shear inference.
\end{itemize}
With such an algorithm to cluster the features in the galaxy population we are 
able to simultaneously fit different distributions for each galaxy 
and exploit the statistical information from a large sample of galaxies. 
As we will see in \autoref{sec:part2}, the DP model (rather than 
a less flexible asserted prior distribution) can be an important component of our framework for improving 
the accuracy and precision of the shear inference.

For later Gibbs sampling we found it more helpful 
to consider the DP model as a finite mixture model 
with $\numclasses$ components in the limit that $\numclasses\rightarrow\infty$~\citep{neal00}.
Let $c_n$ denote the latent class assignment to the mixture that generated 
galaxy $n$. And let $\alpha_{c_n} \equiv \alpha^{*}_{c}$ denote the parameters specifying the 
distribution of $\galprops_n$ for latent class $c_n$. 
Then, our distribution of the intrinsic galaxy components for galaxy $n$ is
\begin{equation}\label{eq:dp1}
  \galprops_{n} | c_{n}, \alpha \sim \prf{\galprops_{n}|\alpha_{c_n}}.
\end{equation}
The latent class assignments $c_n$ are distributed as~\citep[][Eqn. 2.3]{neal00},
\begin{equation}\label{eq:dp2}
  c_n | \pv \sim {\rm Discrete}\left(p_1,\dots,p_{\numclasses}\right),
\end{equation}
for some mixing probabilities $\pv$,
\begin{equation}\label{eq:dp3}
  \pv \sim {\rm Dirichlet}\left(\dpprec / \numclasses, \dots, \dpprec/\numclasses\right),
\end{equation}
with the clustering parameter $\dpprec$ still the same for all mixture components.
The DP model specification is completed with the base distribution $G_0(\alpha_{c})$ defining the 
parameters $\alpha_{c}$ for each each mixture component.
The limit $\numclasses\rightarrow\infty$ of Eqs.~\ref{eq:dp1}--\ref{eq:dp3} is only one of many possible 
ways to represent a DP~\citep{neal00}, but is a convenient representation for our 
Monte Carlo sampling described below.

In the infinite mixture limit, Monte Carlo sampling proceeds by only keeping track of those 
mixture parameters $\alpha_c$ currently associated with an observed galaxy.
Algorithm~2 of \citet{neal00} showed that the conditional probabilities for the latent 
class assignment $c_n$ for observed galaxy $n$ is~\citep[][eq. 3.6]{neal00},
\begin{align}
  \pr(c_n = c_{\ell} | c_{-n}, \galprops_n, \alpha, \condvar) &= b\, N_{-n,c}\, 
  \pr(\data_n | \alpha_{c_{\ell}}, \condvar),
  \notag\\
  &\qquad \forall \ell\neq n
  \label{eq:latent_class_cond_1}
  \\
  \pr(c_n \neq c_{\ell} \forall \ell\neq n | c_{-n}, \galprops_n, \condvar) &=
  \notag\\
  b\, \dpprec \int & \pr(\data_n | \alpha, \condvar)\, G_{0}(\alpha)\, d\alpha,
  \label{eq:latent_class_cond_2}
\end{align}
where $b$ is chosen so the probabilities sum to one,
$N_{-n,c}$ is the number of galaxies currently assigned to latent class $c$ excluding 
galaxy $n$,
$\pr(\data_n | \alpha, \condvar)$ is defined as in \autoref{eq:marg_like},
and $\condvar$ denotes the set of conditional dependencies ($\lenspot, \psf, I$). 
Note Equations \ref{eq:latent_class_cond_1} and \ref{eq:latent_class_cond_2} are similar to Equation \ref{eq:dp_cond_prior_v2} with added information from $\pr(\data_n | \alpha, \condvar)$.
We use Equations \ref{eq:latent_class_cond_1} and \ref{eq:latent_class_cond_2}
 to perform 
Gibbs updates~\citep{tanner96} of the latent class assignment variables $c_n$ 
\citep[see][for more details, but note the mixture probabilites $\pv$ are now integrated out]{neal00}.
Given an update for $\cv\equiv\left\{c_1,\dots,c_n\right\}$, we draw Gibbs updates 
for the mixture parameters as,
\begin{equation}
  \alpha_{c_n} \sim G_0\left(\alpha_{c_n}\right) 
  \prod_{\ell=1}^{N_{c_n}} \pr(\data_{\ell} | \alpha_{c_n}, \condvar),
\end{equation}
where $N_{c_n}$ denotes the number of galaxies associated with latent class $c_n$.
That is, the mixture parameters are drawn from the posterior given all the 
observed galaxies currently associated with mixture class $c_n$ and the prior $G_0(\alpha_{c_n})$.

The ability to use Gibbs sampling for the DP parameters is useful as the number of 
galaxies (and therefore number of DP parameters) becomes large. But, Gibbs sampling 
can also pose challenges both for the flexibility of the statistical model and 
in parallelization of the Monte Carlo sampling routines. We address the former 
challenge in the next sub-section.

\subsubsection{Gibbs updates with importance sampling}
Gibbs sampling for the DP hyperparameters is only practical if the 
integral in \autoref{eq:latent_class_cond_2} can be computed analytically. 
This typically limits the choice of $G_0$ to distributions conjugate to the 
marginal likelihood $\pr(\data_n | \alpha_{c_n}, \condvar)$~\citep[e.g.][]{gorur2010dirichlet}. 
But, \emph{with importance samples for $\galprops_n$, we can perform fast Gibbs updates 
of the DP hyperparameters with the weaker assumption that $G_0$ is 
conjugate to $\prf{\galprops_n | \alpha_{c_n}}$.} That is, 
we are free to specify a DP base distribution without reference to the 
functional dependence of the model parameters $\galprops_n$ in the likelihood.
This is a significant computational advantage for our problem.

For example, in \autoref{sec:numerical_example} we let $\galprops_n$ be 
Gaussian distributed with $\alpha$ a variance parameter. Then we require 
$G_0$ to be conjugate to the variance of a Gaussian distribution. 
If, as in \autoref{sec:numerical_example}, $\galprops_n$ represents 
the galaxy ellipticity, then the likelihood function of the pixel data will 
depend on a nonlinear function of $\galprops_n$ since the ellipticity is a nonlinear 
transformation of pixel values. Already in this idealized scenario it is 
impossible to specify an analytic $G_0$ that is conjugate to the marginal 
likelihood $\pr(\data_n | \alpha_{c_n}, \condvar)$. 
This illustrates how importance sampling is critical
to the success of the DP model specification for the shear inference. 
In \autoref{sub:base_dist_shear_var} we explain how to achieve a wide 
variety of DP base distributions that are conjugate to the 
conditional distribution of intrinsic galaxy properties $\pr(\galprops_n | \alpha_{c_n})$.

Applying the importance sampling formula in \autoref{eq:importance_samples}, 
the marginal likelihood in \autoref{eq:latent_class_cond_1} is
\begin{equation}
  \prf{\data_n | \alpha_{c_l}, \condvar} = \frac{Z_{n}}{N}
  \sum_{k=1}^{N} \frac{\prf{\galprops_{nk} | \alpha_{c_l}}}{\prf{\galprops_{nk} | I_0}}.
\end{equation}
Similarly, for the integral in \autoref{eq:latent_class_cond_2},
\begin{equation}
  \int \pr(\data_n | \alpha, \condvar)\, G_{0}(\alpha|a_{\eta})\, d\alpha =
  \frac{Z_n}{N} \sum_{k=1}^{N}
  \frac{\pr_{\rm marg}(\galprops_{nk} | a_{\eta})}{\pr(\galprops_{nk} | I_0)}
\end{equation}
where,
\begin{equation}\label{eq:marg_prior}
  \pr_{\rm marg}(\galprops_{nk} | a_{\eta}) \equiv \int d\alpha_{c_n}\, G_0(\alpha_{c_n} | a)
  \pr(\galprops_{nk} | \alpha_{c_n}),
\end{equation}
is the distribution for $\galprops_n$ marginalized over the conditional 
PDF parameters $\alpha$ with 
parameters $a$ specifying the form of the DP base distribution $G_0$.
Again, \emph{the only practical requirement on the DP base distribution $G_0$ is that we
be able to compute \autoref{eq:marg_prior} analytically or via fast numerical approximations.}
Recall that $\pr(\galprops_{n}|\alpha_{c_n})$ is the distribution of the intrinsic 
properties of galaxy $n$ given hyperparameters $\alpha_{c_n}$. We can then
restate our requirement on $G_0$ as a requirement for conjugate priors on the 
the intrinsic galaxy properties (such as size, flux, ellipticity, and intensity profile slope).

We describe our implementation of a distribution for $G_0$ in \autoref{sub:base_dist_shear_var}.
In summary we find that by introducing an auxiliary variable $\xi_n$ such that
$\galprops_n \equiv \eta_n \xi_n$, we can allow for flexible, but conditionally conjugate 
distributions for the DP base distribution $G_0$. By allowing the conditional PDF 
on $\galprops_n$ to 
be multivariate Gaussian, we can also accommodate correlations in the parameters of a 
galaxy model.

In \autoref{sec:sampling_dp_clustering_parameter} we describe our sampling model for the 
DP clustering parameter $\dpprec$ given some expectation for the number of distinct 
clusters, or categories, of galaxies in the data.


\subsection{Example inferences of intrinsic source properties}
\label{sec:numerical_example}
\newcommand{\obsvar}{\sigma_{\rm obs}^2}
In this section we use toy models to demonstrate the performance 
of 
1) importance sampling for hierarchical inference of the intrinsic galaxy ellipticity distribution
and 
2) the DP model as a flexible approach to modeling the distributions of galaxy 
intrinsic properties.
Throughout this section we make the strong assumptions that i) the PSF is known 
and ii) the shear is constant among all `observed' galaxies.
These assumptions are not realistic but are intended to simplify the demonstration 
of important numerical features of our statistical framework.

\begin{figure*}[htb]
  \centerline{
    \includegraphics[width=0.4\textwidth]{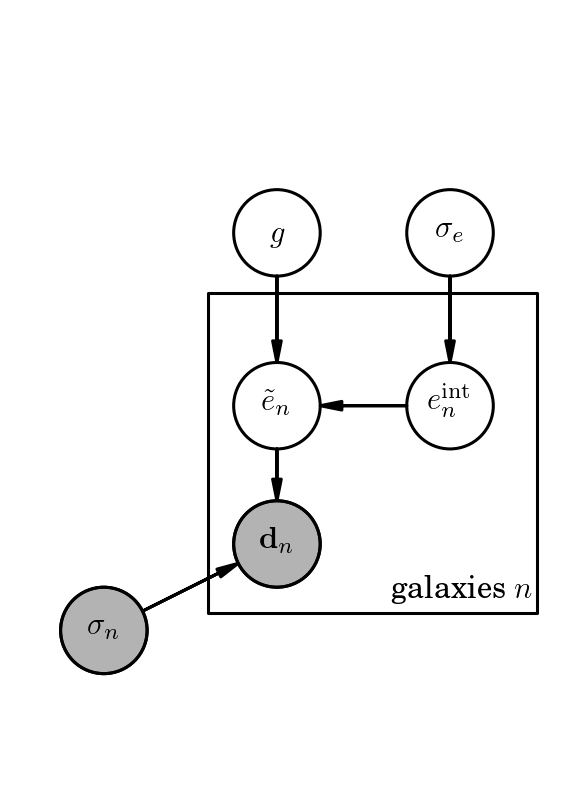}
    \includegraphics[width=0.4\textwidth]{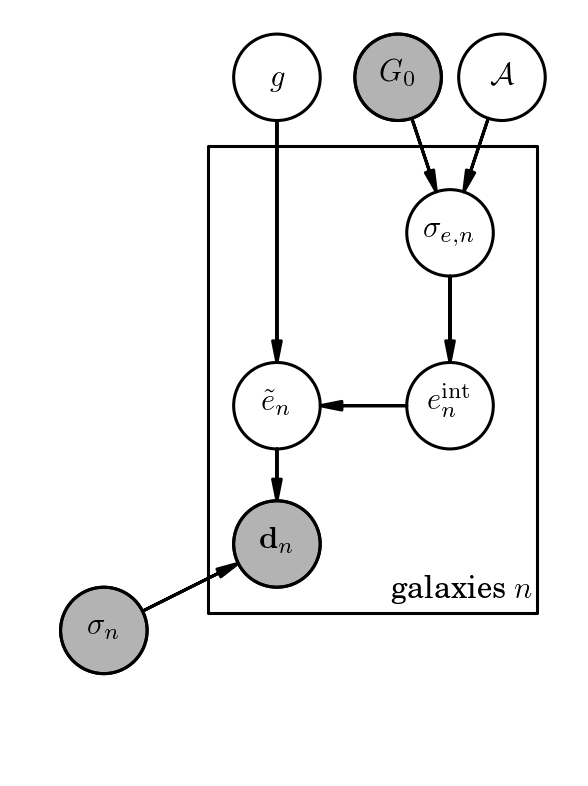}    
  }
  \caption{Probabilistic graphical models for the two toy models 
  in \autoref{sec:numerical_example}.
  Left: Toy model 1 (\autoref{sub:relative_gains_using_importance_sampling})
  is set up to illustrate how marginalizing over $\sigma_{e}$ reduces biases 
  in the inferred shear $g$.
  Right: Toy model 2 (\autoref{sub:dp_inference}) is set up to illustrate how 
  marginalizing over possible distributions for $\sigma_{e}$ improves the 
  marginal shear inference in cases where the intrinsic ellipticity is 
  generated from multiple statistical populations.
  }
  \label{fig:toy_model_pgms}
\end{figure*}

For our toy models we introduce the following notation for ellipticities,
$\shapeint$ is intrinsic ellipticity in a galaxy model, 
$\elensed$ is the galaxy model ellipticity after applying a shear $g$ to the model,
$\shapeobs$ is an estimator for ellipticity computed from observed pixel values.

\subsubsection{Toy model 1: Benefits of hierarchical inference} 
\label{sub:relative_gains_using_importance_sampling}

Marginalizing the parameters of the intrinsic galaxy ellipticity distribution 
can reduce biases in the shear inferences in some cases. 
We illustrate this point with a simple statistical model that 
we will refer to as `toy model 1'.

We construct a simplified model for the shear inference with the following 
assumptions:
\begin{enumerate}
  \item the pixel data for each galaxy is reduced to `observed' ellipticity components $\shapeobs$
  with observational uncertainties described by uncorrelated Gaussian noise with variance 
  $\obsvar$ for all galaxies,
  \item the intrinsic galaxy ellipticity $\shapeint$ is generated from a zero-mean Gaussian 
  distribution and is inferred with a similar hyper-prior,
  \item we can work in the weak shear regime where $\elensed\approx \shapeint + g$.
\end{enumerate}
We illustrate the dependencies of toy model 1 in the left panel of \autoref{fig:toy_model_pgms}.
This model is simple enough that we can derive the marginal 
posteriors for both the shear and the intrinsic ellipticity variance analytically. 
We perform these derivations in the next subsection and then use the 
results to build some intuition about the benefits of hierarchical inference of 
distributions of intrinsic galaxy properties in 
\autoref{sub:shear_maximum_a_posteriori_estimators}.

\paragraph{Marginal posteriors for toy model 1}
Because the intrinsic (i.e., unsheared) galaxy shapes are always unknown, 
we first marginalize over $\shapeint$ in any analysis.
For this toy model, 
marginalizing over the intrinsic ellipticities for each galaxy yields a Gaussian 
marginal likelihood for the observed ellipticities of each galaxy given a 
common variance $\sigma_e^2$ for the intrinsic galaxy ellipticities,
\begin{align}\label{eq:tm1_marg_like}
  \pr(\shapeobs | g, \sigma_e^2) &\propto
  \notag\\
  \int d^2\shapeint &
  \normdist_{\shapeobs} \left(\shapeint+g, \obsvar\right)
  \cdot
  \normdist_{\shapeint}\left(0, \sigma_{e}^2\right)
  \notag\\
  &= \normdist_{\shapeobs} \left(g, \obsvar + \sigma_{e}^{2}\right).
\end{align}
The marginal likelihood for $\ngal$ galaxies is just the product of $\ngal$ 
versions of \autoref{eq:tm1_marg_like},
\begin{align}
  \pr\left(\left\{\shapeobs_{i}\right\}_{i=1}^{\ngal} | g, \sigma_e^2\right)
  &\propto
  \prod_{i=1}^{\ngal} \pr(\shapeobs_{i} | g, \sigma_e^2)
  \notag \\
  = \left(2\pi \sigma^2\right)^{-\ngal/2}
  &\exp \left[-\frac{\ngal}{2\sigma^2}{\rm Var}\left\{\shapeobs_{i}\right\}\right]
  \notag\\
  &\times\exp \left[-\frac{\ngal}{2\sigma^2} \left|g-\shapeobsmean\right|^2\right],
\end{align}
where we defined,
\begin{equation}
  \shapeobsmean \equiv \frac{1}{\ngal}\sum_{i=1}^{\ngal} \shapeobs_{i},
  \quad {\rm and}\quad
  \sigma^2 \equiv \obsvar + \sigma_{e}^{2}.
\end{equation}

The conditional posterior for the shear $g$ is a Gaussian distribution
given $\ngal$ galaxy observations and a known (or assumed) value for the
intrinsic ellipticity variance $\sigma_e^2$,
\begin{align}\label{eq:tm1_shear_cond_post}
  \prf{g | \sigma_e^2, \left\{\shapeobs_{i}\right\}}
  &\propto
  \pr\left(\left\{\shapeobs_{i}\right\} | g, \sigma_e^2\right)
  \cdot
  \normdist_{g} \left(0, \sigma_{g}^2 \ident_{2}\right)
  \notag\\
  &\propto \normdist\left(\mu_{cg}, \sigma^2_{cg}\ident_2\right)
\end{align}
with,
\begin{align}
  \mu_{cg} &\equiv \shapeobsmean \left(1 + \frac{\sigma^2}{\ngal \sigma_{g}^2}\right)^{-1}
  \label{eq:tm1_shear_cond_mean}
  \\
  \sigma^2_{cg} &\equiv \frac{\sigma^2 / \ngal}{1 + \frac{\sigma^2}{\ngal\sigma^2_g}}.
\end{align}
\autoref{eq:tm1_shear_cond_post} represents the shear posterior that is 
realized in any analysis that does not attempt to simultaneously infer the 
intrinsic ellipticity distribution. It is useful to consider 
some limiting cases of \autoref{eq:tm1_shear_cond_post}.

\autoref{eq:tm1_shear_cond_mean} suggests defining the composite variable,
\begin{equation}
  x \equiv \frac{\sigma^2}{\ngal\sigma_g^2}.
\end{equation}
Then,
\begin{equation}
  \mu_{cg} = \frac{\shapeobsmean}{1+x},\qquad
  \sigma^2_{cg} = \sigma_g^2 \frac{x}{1+x}.
\end{equation}
So, point estimators for $g$ conditioned on $\sigma_e$ will become 
more biased towards zero as $x$ increases (e.g., for small numbers of galaxies, 
an informative zero-mean shear prior, or a large dispersion in the intrinsic galaxy ellipticities).
The variance on the maximum posterior estimator for $g$ becomes arbitrarily small as $x\rightarrow0$ 
(e.g., as $\ngal\rightarrow\infty$) but is bounded from above by the value of 
$\sigma_g^2$ as $x$ becomes large.

Marginalizing over $\sigma_e^2$ in \autoref{eq:tm1_marg_like} 
requires performing the integral,
\begin{multline}\label{eq:shear_marg_posterior}
  \int_{0}^{\infty} dx\, (\obsvar + x)^{b} \exp \left(-\frac{a}{\obsvar+x}\right)
  \\
  = a^{1-b}\gamma\left(b-1, \frac{a}{\obsvar}\right),
\end{multline}
where $\gamma(b-1, a/\obsvar)$ is the lower incomplete Gamma function,
and,
\begin{align}
  b &\equiv \ngal \notag\\
  a &\equiv \ngal {\rm Var}\left\{\shapeobs_{i}\right\} + 
  \ngal \half \left|g-\shapeobsmean\right|^2.
\end{align}

Marginalizing over the shear gives the following posterior distribution 
for the intrinsic ellipticity variance,
\begin{multline}\label{eq:sigmaesq_marg_posterior}
  \pr\left(\sigma_{e}^{2} | \left\{\shapeobs_{i}\right\}, \sigma_g^2\right)
  \\
  \propto
  \Gamma^{-1} \left(\ngal-2, \ngal\half{\rm Var}\left\{\shapeobs_{i}\right\}\right)
  \\
  \times
  \normdist_{\shapeobsmean} \left(0, \sigma_{g}^{2} + \frac{\sigma^2}{\ngal}\right),
\end{multline}
where $\Gamma^{-1}$ is the inverse-Gamma distribution.

\begin{figure*}[htb]
  \centerline{
    \includegraphics[width=0.5\textwidth]{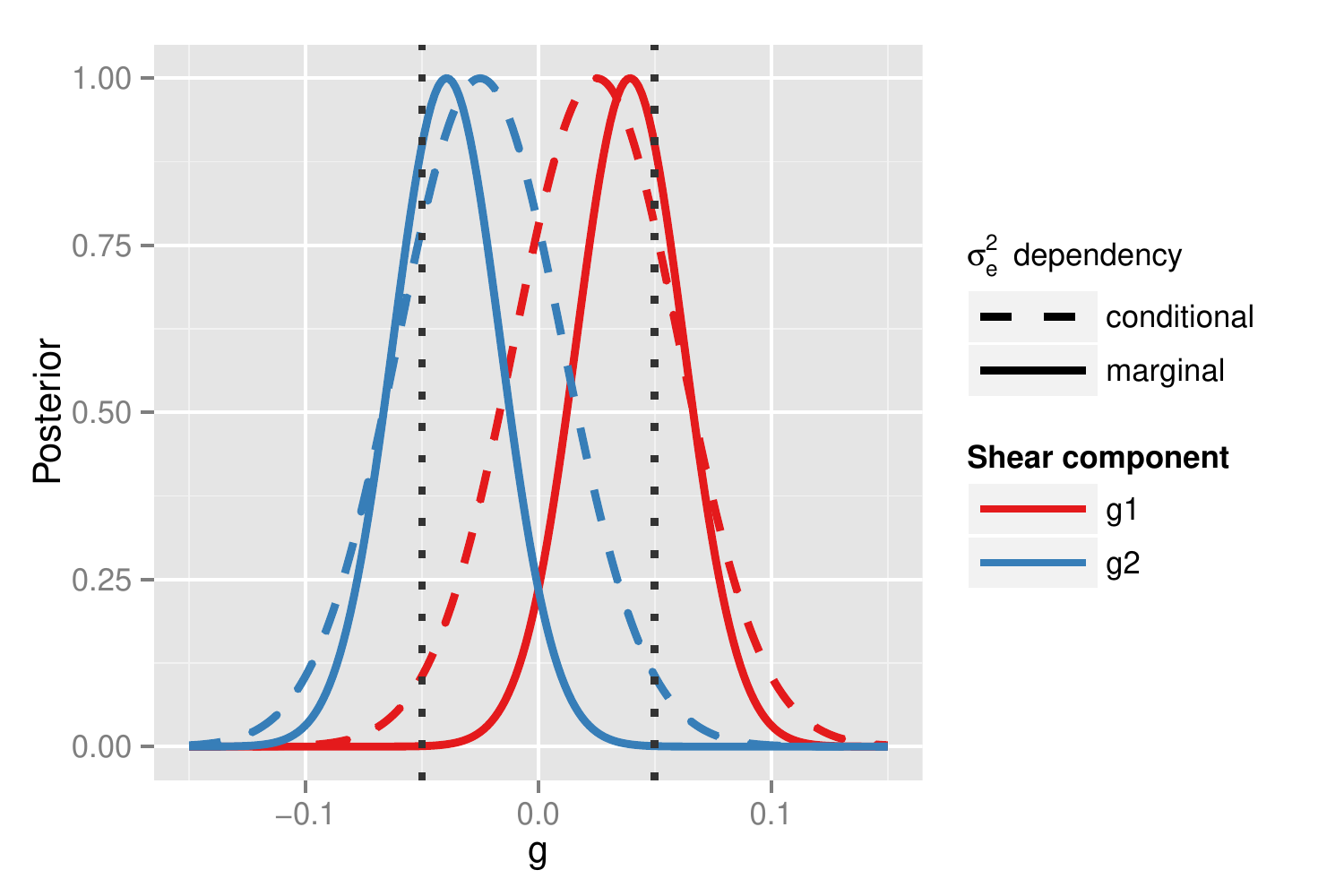}
    \includegraphics[width=0.335\textwidth]{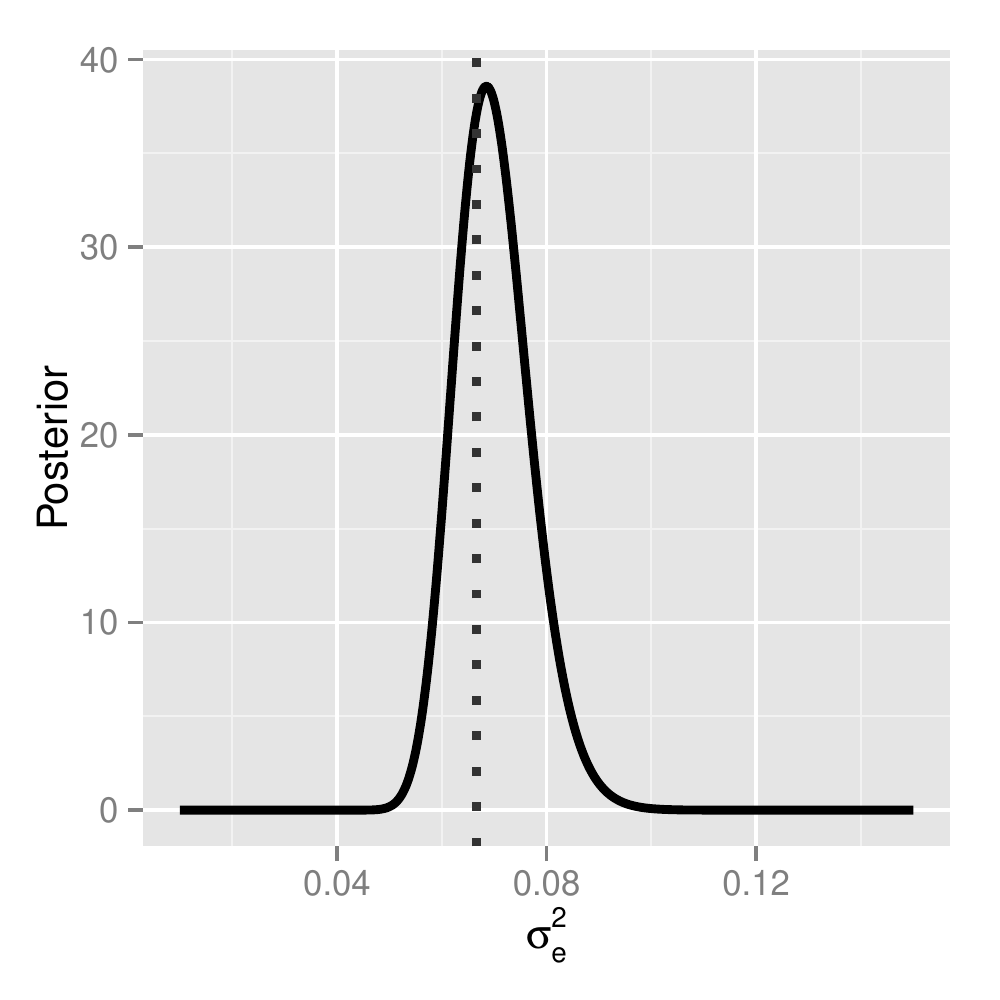}    
  }
  \caption{Conditional and marginal posteriors for the shear and intrinsic ellipticity variance 
  in toy model 1. 
  The vertical dotted lines denote the values used to generate the data.
  See \autoref{tab:tm1_parameters} for the parameter values used in this
  figure.}
  \label{fig:tm1_marg_posteriors}
\end{figure*}

We plot the expressions in Equations~\ref{eq:tm1_shear_cond_post}, 
\ref{eq:shear_marg_posterior}, and \ref{eq:sigmaesq_marg_posterior} 
for the toy model 1 conditional and marginal posteriors 
in \autoref{fig:tm1_marg_posteriors}.
We have selected parameters for \autoref{fig:tm1_marg_posteriors}, listed in 
\autoref{tab:tm1_parameters}, that 
illustrate how a large bias in the shear from the posterior conditioned on 
$\sigma_e^2$ can be alleviated by marginalizing $\sigma_e^2$. 
However, many combinations of 
parameters in toy model 1 will result in smaller shear biases.

\begin{table}[htb]
\begin{center}
\caption{Parameters for toy model 1 posterior plots}
\label{tab:tm1_parameters}
\begin{tabular}{cc}
\hline \hline
Parameter & Value \\
\hline
$g_{1,2}$ & (0.05, -0.05) \\
$\obsvar$ & $7.6\times 10^{-5}$ \\
$\ngal$ & 100 \\
$\sigma_g$ & 0.05 \\
$\sigma_e$ & $0.258$ \\
bias in assumed $\sigma_e$ & 0.242 \\
\hline\hline
\end{tabular}
\end{center}
\end{table}

The value of $\sigma_e$ in \autoref{tab:tm1_parameters} is derived from a Gaussian 
fit to the distribution of ellipticity values in the Deep Lens Survey (DLS)\footnote{\url{http://matilda.physics.ucdavis.edu/working/website/index.html}}~\citep{2002SPIE.4836...73W,jee2013} 
(with shear included, but after correction for the PSF).
We assume 100 galaxies in \autoref{tab:tm1_parameters} 
both because the shear may not be expected to be constant over a 
sky area containing larger numbers of galaxies, and because the GNU~Scientific~Library\footnote{\url{http://www.gnu.org/software/gsl/}}~\citep{gsl} 
routine we use to evaluate \autoref{eq:shear_marg_posterior} does not yield numerically stable 
results for larger $\ngal$ values.

The final line of \autoref{tab:tm1_parameters} indicates the `bias' in the 
value of $\sigma_e$ that is assumed in the conditional posterior in \autoref{eq:tm1_shear_cond_post}, 
relative to the true value of $\sigma_e=0.258$. With this bias of $0.242$, 
the assumed value of $\sigma_e^{\rm assumed}=0.5$, indicating a value that might be chosen as 
`non-informative' in a non-hierarchical analysis. \autoref{fig:tm1_marg_posteriors} then 
shows that knowledge of the true distribution of intrinsic ellipticities (via hierarchical inference) can be important 
in mitigating shear biases and it is not sufficient to assert a broad intrinsic ellipticity 
prior.

\paragraph{Prior choices for shear point estimators}
\label{sub:shear_maximum_a_posteriori_estimators}

Most published cosmic shear analyses rely on point estimators for the 
shear of each galaxy image. 
We do not advocate the use of point estimators for the shear. 
But in this subsection we consider the shear posterior mean as a 
possible point estimator that might be compared with other point estimators 
in the literature.

The mean of the shear conditional posterior, \autoref{eq:tm1_shear_cond_post},
is least biased relative to the true shear 
when $\sigma_g$ is large and the true value of $\sigma_{e}$ is known.
The mean of the marginal posterior for the shear, \autoref{eq:shear_marg_posterior}, does not require knowledge 
of $\sigma_{e}$ since it has been integrated out and is similarly least biased 
when $\sigma_g$ is large. These statements of course depend on our assumption of a 
zero-mean shear prior.

\begin{figure*}[htb]
  \centerline{
		\includegraphics[width=\textwidth]{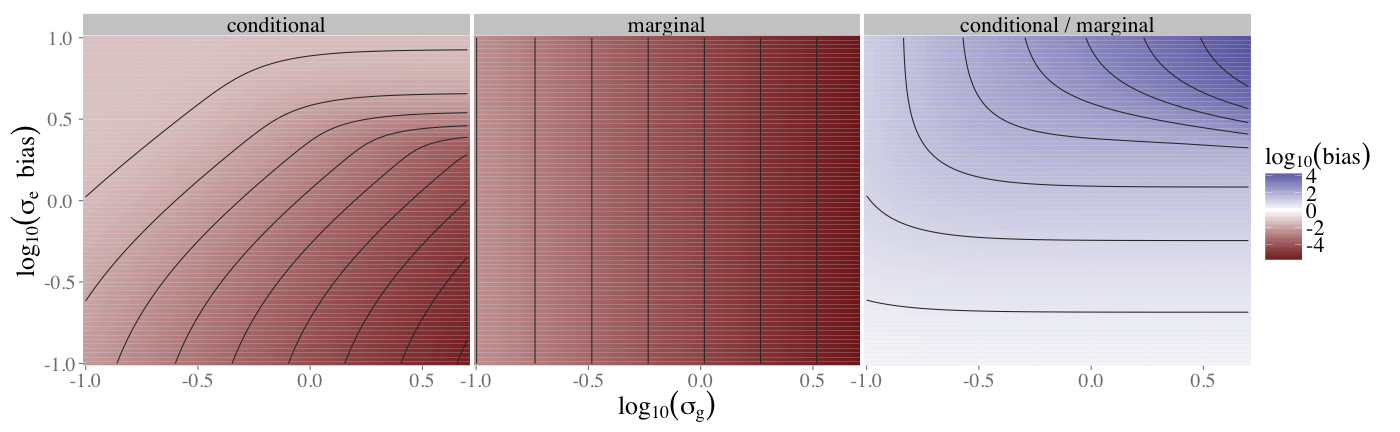}
	}
	\caption{Biases in the posterior means for a component of the shear.
  The left panel shows the posterior mean bias when conditioned on the 
  assumed width of the intrinsic ellipticity distribution,
  $\sigma_{e}$.
  The central panel shows that of the posterior marginalized over $\sigma_{e}$ (which is 
  insensitive to biases in assumed $\sigma_{e}$).
  The right panel is the ratio of the shear biases in the left and central panels.
  The vertical axis shows the logarithm of the additive bias in the assumed $\sigma_{e}$ relative 
  to the true value (so a perfect guess for the true width of the intrinsic ellipticity distribution 
  corresponds to $-\infty$ on the vertical axis). 
  The horizontal axis shows the logarithm of the width of the zero-mean prior for the shear components.
  For all assumed values of $\sigma_e$ and $\sigma_g$, marginalizing over $\sigma_e$ always gives 
  less biased shear posterior means (comparing the central to the left panel). 
  In addition, assuming broad priors (i.e., weakly informative) for the intrinsic ellipticity and the 
  shear gives the most biased shear inferences if one does not marginalize over the 
  intrinsic ellipticity distribution width (right panel).
  }
	\label{fig:tm1_shear_bias}
\end{figure*}

In \autoref{fig:tm1_shear_bias} we show the bias in the mean of the 
conditional shear posterior and in the marginal shear posterior as a function 
of $\sigma_g$ and the bias in the assumed value of $\sigma_e$ relative to the 
value used to generate the data. We can see that the \emph{relative} bias 
(right panel of \autoref{fig:tm1_shear_bias}) in the mean 
between the conditional and marginal shear posteriors is largest for large 
$\sigma_g$ and large assumed $\sigma_e$.
One might have na\"{i}vely guessed that flat priors in shear and intrinsic ellipticity 
(equivalent in our toy model to large $\sigma_e$ and $\sigma_g$) would be the obvious 
choice for shear inference in the absence of any other information. 
But, \autoref{fig:tm1_shear_bias} illustrates that such flat priors will give the 
most biased shear inference relative to what could be obtained with a hierarchical model
for the intrinsic galaxy ellipticities.

\subsubsection{Toy model 2: DP inference given a bimodal intrinsic ellipticity distribution} 
\label{sub:dp_inference}

We now construct a numerical toy model to illustrate how the marginal shear inference can 
be improved by taking advantage of the flexibility of the DP as a conditional PDF for the distribution of intrinsic 
galaxy properties instead of asserting a simple Gaussian distribution for the galaxy properties.
We will refer to the models in this subsection collectively as `toy model 2'.

As in toy model 1, we assume the pixel data is reduced to observed ellipticities. 
Unlike in toy model 1 we no longer assume weak shear or a fixed correspondence between 
the intrinsic ellipticity generating distribution and PDF for inference. 

Our objective with toy model 2 
is to compare the marginal shear and ellipticity inferences when using either a
restrictive or flexible conditional PDF for $\sigma_e^2$, irregardless of the generating distribution
for the data.

For a restrictive PDF, we specify a uniform prior on a single $\ln(\sigma_e)$
parameter for all galaxies, effectively assuming that all galaxy ellipticities are drawn from a common 
zero-mean Gaussian distribution. We call this PDF choice our `Gaussian' model.
We use the DP model for $\sigma_e^2$ as a flexible hierarchical 
PDF specification, which we label as our `DP' model.
We describe each model in \autoref{tab:tm2_models}.
\begin{table}[htb]
\begin{center}
\caption{Intrinsic ellipticity PDFs for toy model 2.}
\label{tab:tm2_models}
\begin{tabular}{ccc}
\hline \hline
Model & $\pr(\shapeint_{i})$ & $\pr(\sigma_e^2)$ \\
\hline
Gaussian     & $N\left(0, \sigma_{e}\right)$   & $1 / \sigma_{e}^{2}$     \\
DP     & $N\left(0, \sigma_{e,i}\right)$ & $\dirproc(\dpprec, G_{0})$ \\
\hline \hline
\end{tabular}
\end{center}
\end{table}
\autoref{fig:toy_model_pgms} shows the relationships of the parameters in the Gaussian and DP models
(left and right panels). 

To test and compare the `Gaussian' and `DP' models (distinct from the prescription for generating data) 
we generate two mock data sets, each with 100 galaxies.
For each data set we draw samples of $\shapeint$ from a two-component Gaussian mixture,
\begin{equation}
  \pr(\shapeint) = \lambda \normdist(0, \sigma_{e,1})+ (1-\lambda) \normdist(0, \sigma_{e,2}),
\end{equation}
with parameters given in \autoref{tab:tm2_data}.
\begin{table}[htb]
\begin{center}
\caption{Mock data parameters for toy model 2.}
\label{tab:tm2_data}
\begin{tabular}{cccc}
\hline \hline
Data set & $\sigma_{e,1}$ & $\sigma_{e,2}$ & mixture fraction $\lambda$ \\
\hline
Unimodal     & 0.258 & 0    & 1   \\
Bimodal      & 0.258 & 0.03 & 0.7 \\
\hline \hline
\end{tabular}
\end{center}
\end{table}
The `unimodal' data set has $\lambda = 1$ indicating the $\shapeint$ samples are drawn from a single 
Gaussian distribution (matching the distribution in toy model 1). For the `bimodal' data set, 
we draw $\shapeint$ from a bimodal distribution with 70\% of galaxy ellipticities 
sampled from the Gaussian with width matching that of the unimodal data set, and 30\% sampled from a 
much narrower distribution of intrinsic ellipticity. 
We describe our detailed procedure for generating mock data sets in Appendix~\ref{sec:fake_reaper}.
\autoref{fig:fake_reaper_data} shows the observed ellipticity 
components generated for each mock data set.
\begin{figure}[htb]
  \centerline{
    \includegraphics[width=0.5\textwidth]{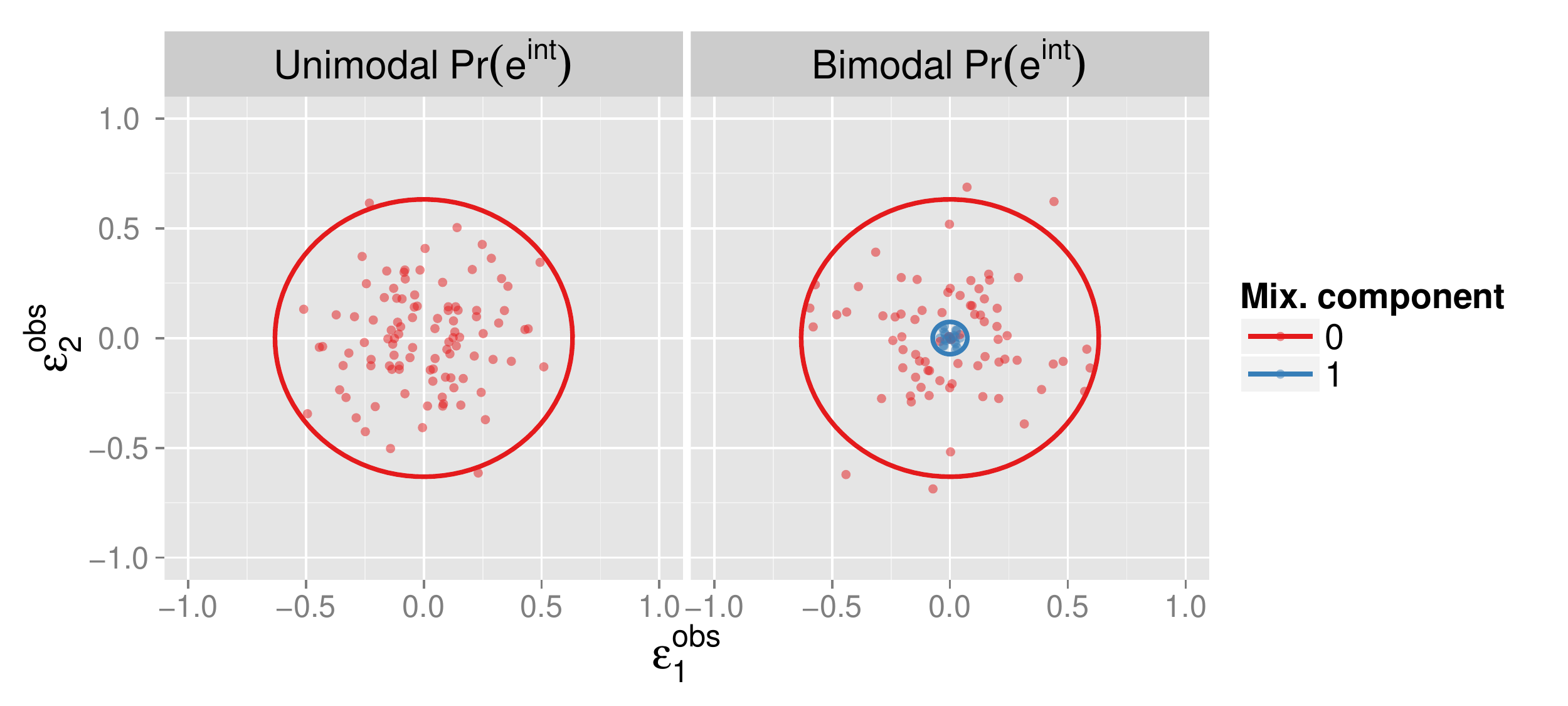}
  }
  \caption{Mock data for 100 simulated galaxy ellipticities for toy model 2.
  The intrinsic ellipticity components are generated from two-component Gaussian mixture 
  models with parameters given in \autoref{tab:tm2_data}.
  }
  \label{fig:fake_reaper_data}
\end{figure}
We motivate the values of $\sigma_{e,1}$ and $\sigma_{e,2}$ in \autoref{tab:tm2_data} by 
fitting one- or two-component Gaussian mixtures to the distribution of observed 
ellipticities from the Deep Lens Survey (DLS)~\citep{jee2013}. These 
ellipticities have been PSF corrected, but include the cosmic shear. We find that 
a two-component Gaussian mixture is a better fit to the observed ellipticity distribution,
giving some motivation for the two-component mixture in our bimodal data set.
However, in \autoref{tab:tm2_data} we have artificially increased the fraction of galaxies 
with a narrow ellipticity distribution over that found in the DLS (where we find a mixture 
fraction of $\sim90\%$ for the broader ellipticity distribution).

\begin{figure}[htb]
  \centerline{
    \includegraphics[width=0.5\textwidth]{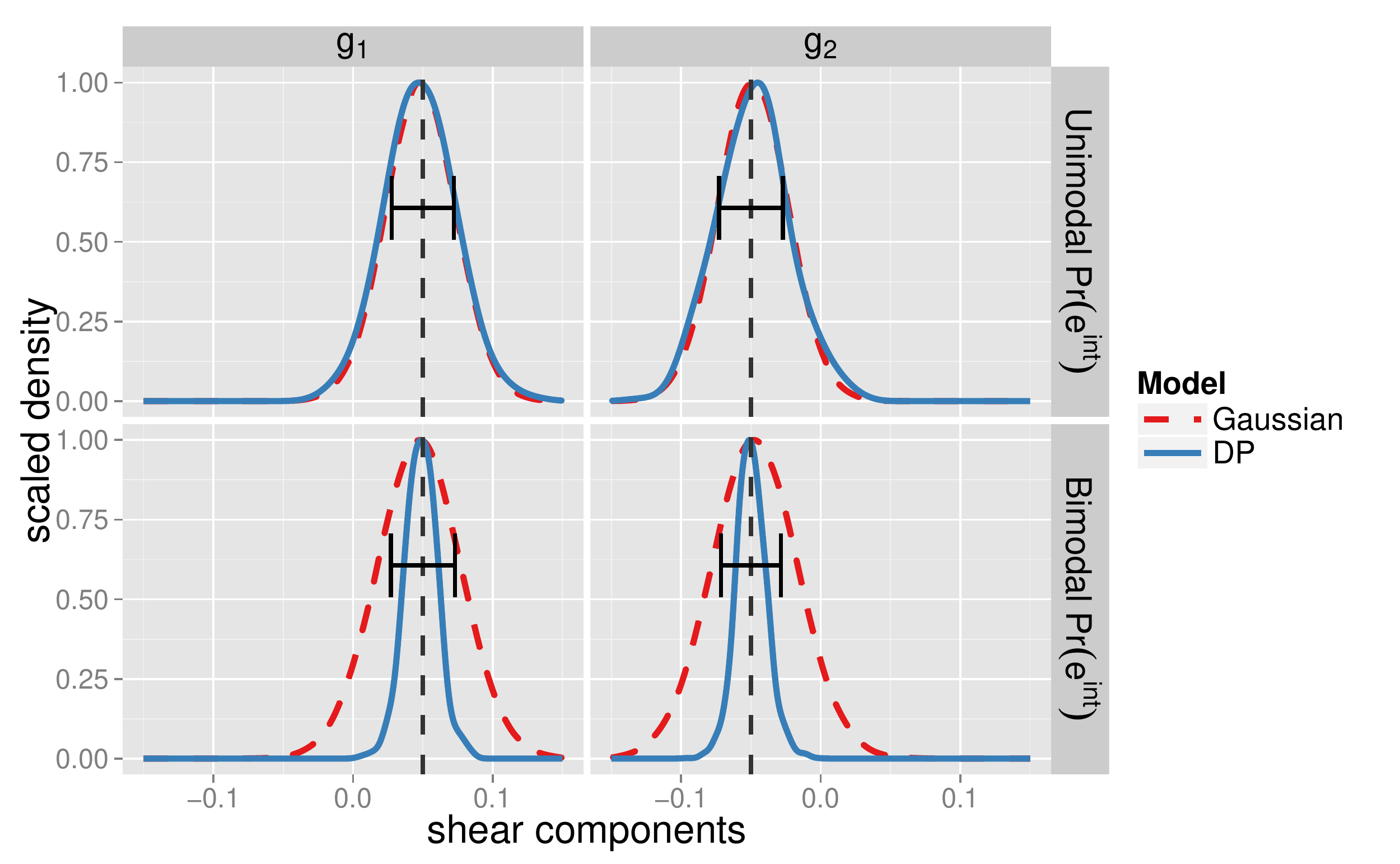}
  }
  \centerline{
    \includegraphics[width=0.5\textwidth]{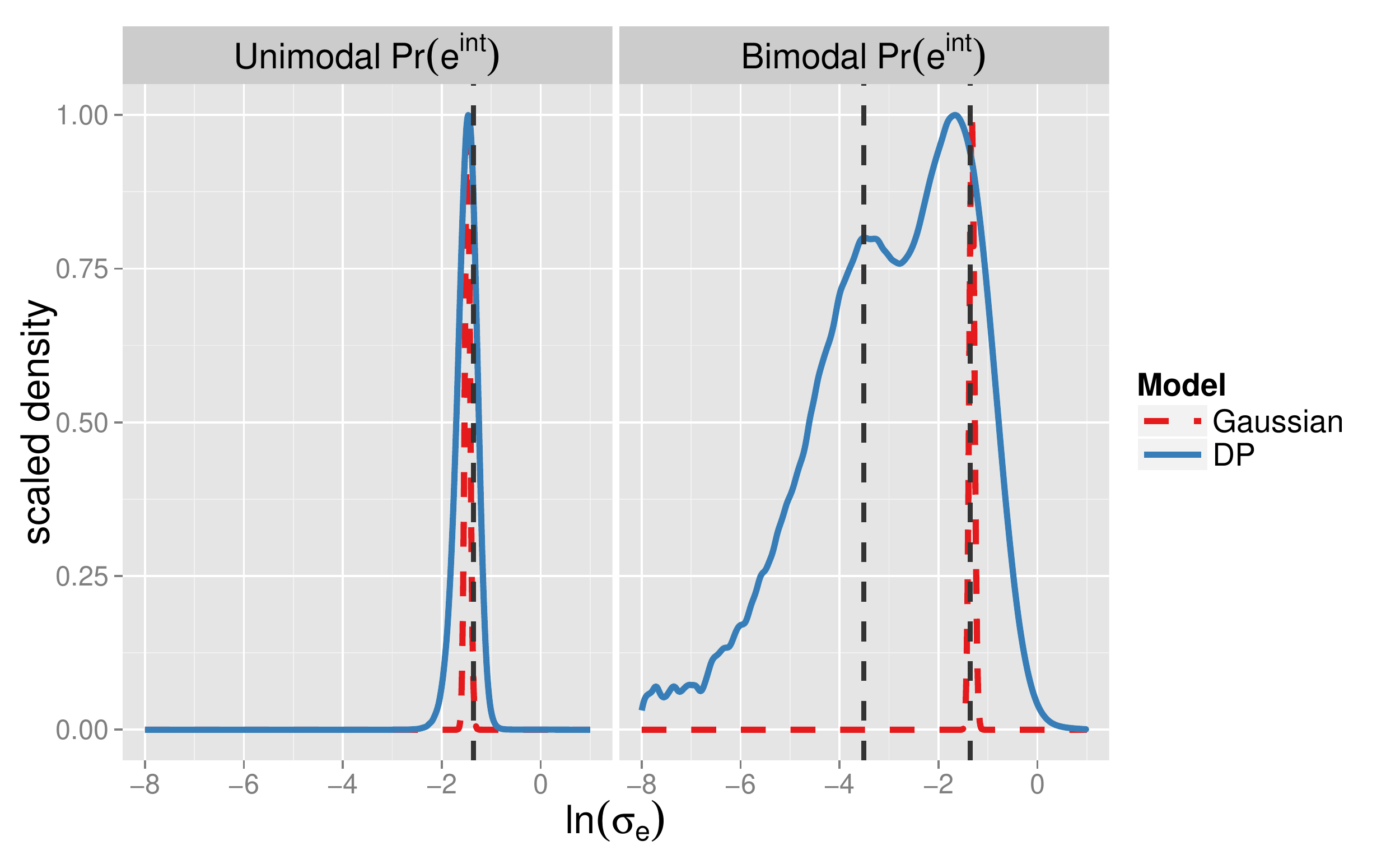}    
  }
  \caption{
  Top: Marginal posteriors for the shear components under the four toy model 2 scenarios.
  The horizontal black lines with error bars indicate the mean and uncertainty on the 
  shear estimator obtained from averaging the observed ellipticity components of all galaxies.
  Bottom: Marginal posteriors for the logarithm of the standard deviation parameter in the 
  intrinsic ellipticity distribution. For the DP model, we compute 
  marginal densities from the concatenation of MCMC samples for $\sigma_{e,i}$
  with $i=1,\dots,\ngal$.
  The vertical dashed lines in all panels show the `true' values corresponding to the circles in \autoref{fig:fake_reaper_data}.
  }
  \label{fig:numerical_example_results}
\end{figure}

\autoref{fig:numerical_example_results} shows the 
marginal posteriors for the shear and the intrinsic ellipticity variance as estimated 
from MCMC samples. 
We describe the conditional distributions for Gibbs sampling the DP model parameters 
for toy model 2 in Appendix~\ref{sec:dp_sampling_algorithm_for_toy_model_2}.
We note here however, that since we have not yet specified an algorithm to determine the 
DP hyperprior parameters $\tau$ and $m$ (see equation~\ref{eq:halfcauchy_prior} and surrounding discussion), 
we have assigned informative priors for these parameters 
that are different for the unimodal and bimodal data sets given our knowledge of the 
generating distributions\footnote{Generally we want to define a DP base distribution $G_0$ with larger probability mass around zero when we expect fewer distinct 
sub-populations in the data. This is non-intuitive given the variable expansion. 
$\sigma_e\equiv |\xi|\sigma_{\eta}$ with $\sigma_{\eta}\sim DP(\dpprec, G_0)$.}. 
An important area for future work will be to dynamically set or sample in the hyperprior parameters for 
the DP base distribution.

The horizontal black lines in the top panel of \autoref{fig:numerical_example_results} indicate the 
uncertainty on the na\"{i}ve shear estimator obtained by averaging the observed ellipticities of the 
galaxies (i.e., the horizontal line gives 
sd$\left(\left\{\shapeobs_{i}\right\}_{i=1}^{\ngal}\right)/\sqrt{\ngal})$.
These na\"{i}ve shear estimator lines are positioned at the hieght of the 1-$\sigma$ width of the 
posterior densities for each model.

\begin{table}[htb]
\begin{center}
\caption{Ratios of mean marginal shear biases using the DP or Gaussian models to 
marginalize the intrinsic ellipticities for toy model 2.}
\label{tab:shear_bias_ratios}
\begin{tabular}{cc}
\hline\hline
Data set & DP model / Gaussian model \\
         & $(g_1, g_2)$ \\
\hline
Unimodal & (1.26, 1.23) \\
Bimodal & (0.46, 0.12) \\
\hline \hline
\end{tabular}
\end{center}
\end{table}

For the marginal shear posteriors with the unimodal data set 
(top row of the top panel in \autoref{fig:numerical_example_results}), both the Gaussian and DP models yield 
similar results, which are similar as well to results obtained using a simple average of the ellipticities to estimate the shear.

However, the shear marginals for the bimodal data set 
(lower row of the top panel of \autoref{fig:numerical_example_results})
show large differences between the Gaussian and DP models. 
The Gaussian model yields shear marginal distributions similar to 
those obtained with the unimodal data set, 
while the DP model yields shear marginals that are more than twice as precise as 
those assuming the Gaussian model 
(and are narrower than the shear errors obtained from averaging ellipticities).
In \autoref{tab:shear_bias_ratios} we give the ratio of the biases in the mean 
$g_1, g_2$ from the DP model to that from the Gaussian model for each of the mock data sets. 
For the bimodal data set, the DP model yields mean shear estimates that are less biased than the Gaussian model, while the 
reverse is true for the unimodal data set. 
These results are better understood by next investigating the $\sigma_e$ marginal distributions 
in the lower panel of \autoref{sec:numerical_example}.

For the unimodal data set, both models yield accurate inferences of $\sigma_e$, although 
the DP model gives a broader marginal posterior than the Gaussian model. However, for the bimodal data set 
(right half of the lower panel of \autoref{sec:numerical_example}, the Gaussian model 
yields a tight posterior for $\sigma_e$ that misses the existence of a sub-population of 
galaxies generated from a much narrower intrinsic  ellipticity distribution.
The DP model yields a bimodal marginal posterior for $\sigma_e$, indicating some knowledge of the 
bimodal nature of the generating distribution.

We can now understand the marginal shear posteriors given the bimodal data set as follows. 
The Gaussian model effectively assumes the entire galaxy population is generated from a broad 
intrinsic ellipticity distribution, which limits the precision with which the shear can 
be measured. The widths of the the Gaussian model shear marginals given the unimodal and 
bimodal data sets are similar 
because the inference assumes similar width ellipticity generating distributions.
However, because the DP model accurately infers the existence of a galaxy sub-population with 
small intrinsic ellipticities, this information can be used to more precisely infer the 
shear for the whole population. 

\begin{figure}[htb]
  \centerline{
    \includegraphics[width=0.5\textwidth]{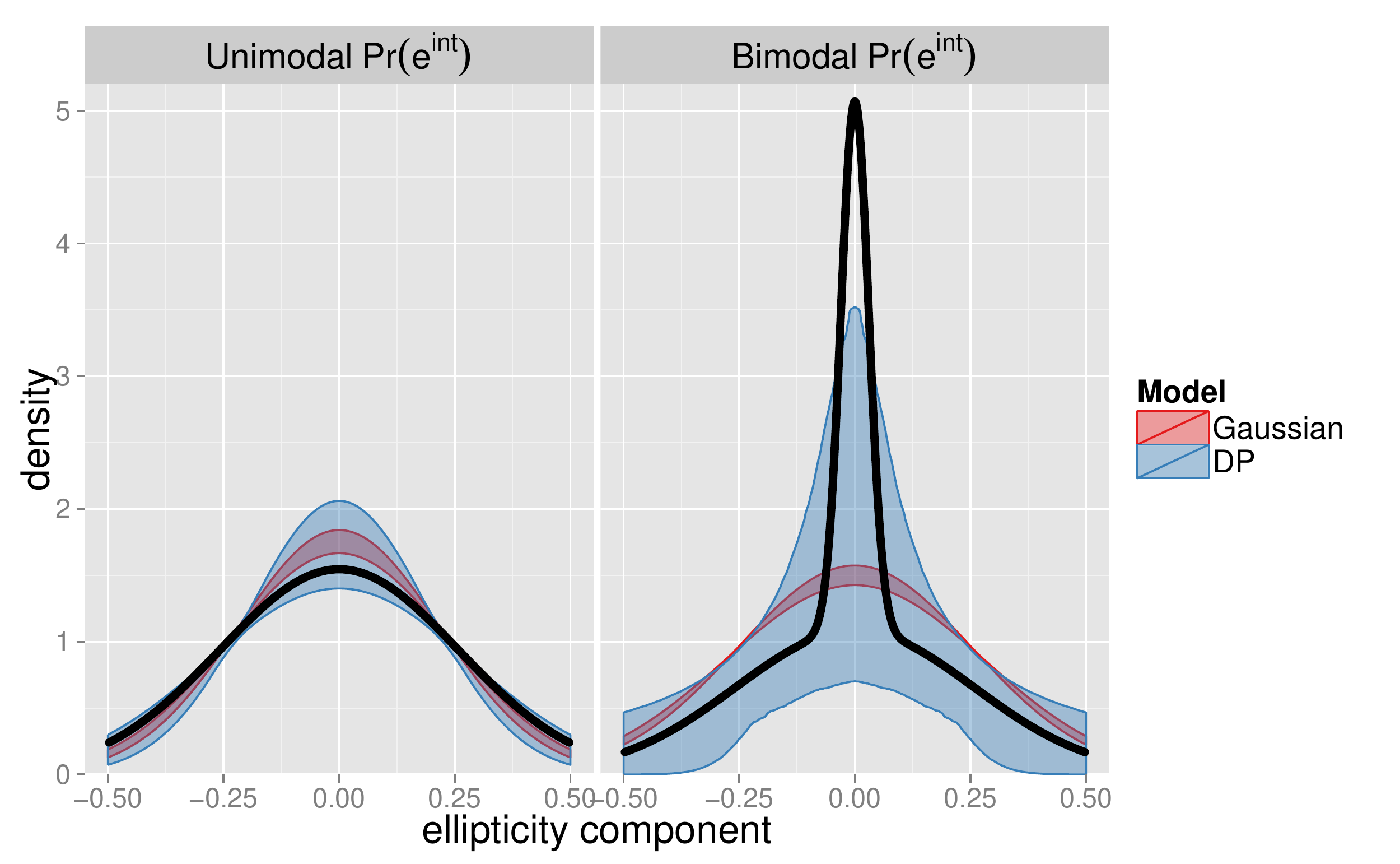}
  }
  \caption{Realizations of the intrinsic ellipticity distribution given marginal 
  posterior samples of $\sigma_e$ in toy model 2.
  The black lines show the generating distributions for mock data scenarios 
  `data1' and `data2'.
  The shaded bands indicate the 68\% credible intervals on the 
  realized ellipticity probability densities.
  In both mock data scenarios, the credible interval for `model1', which 
  assumes $\sigma_e$ is the same for all galaxies, excludes the generating 
  distributions.
  }
  \label{fig:realized_eint_pdfs}
\end{figure}
\autoref{fig:realized_eint_pdfs} shows another way to visualize and interpret the
inferences of the intrinsic ellipticity distributions with both models. 
The solid lines in \autoref{fig:realized_eint_pdfs} show the generating distributions 
of $\shapeint$ for the unimodal and bimodal data sets. The shaded bands show the 68\% credible intervals 
for the inferred $\shapeint$ distributions given marginal posterior samples of 
$\sigma_e$ from both models. While the shaded band for the Gaussian model is not dissimilar 
from the generating distribution for the unimodal data set, it is a poor approximation to the generating 
distribution for the bimodal data set. We should therefore expect larger shear errors and biases when using 
the Gaussian model to analyze the bimodal data set than when using a more flexible 
ellipticity conditional PDF as in the DP model.

In the case of the bimodal generating distribution,
we might obtain more precise and accurate marginal shear inferences from a 
model for the intrinsic ellipticity distribution that is still unimodal, but calibrated from 
a sub-sample of the data to partially include information about a rounder subpopulation of 
galaxies with a different response to shear.
The real power of the DP, however, is that it not only helps to infer the underlying distribution 
of intrinsic ellipticities from the data but that it also enables assigning different 
galaxies to different latent classes. Thus, it is capable of determining which 
galaxies to weight more in the shear measurement (i.e., those that belong to classes with 
smaller intrinsic ellipticity variance). The DP inference thereby results in not only a 
less biased shear inference but also a more precise shear inference.

\begin{figure}[htb]
  \centerline{
    \includegraphics[width=0.5\textwidth]{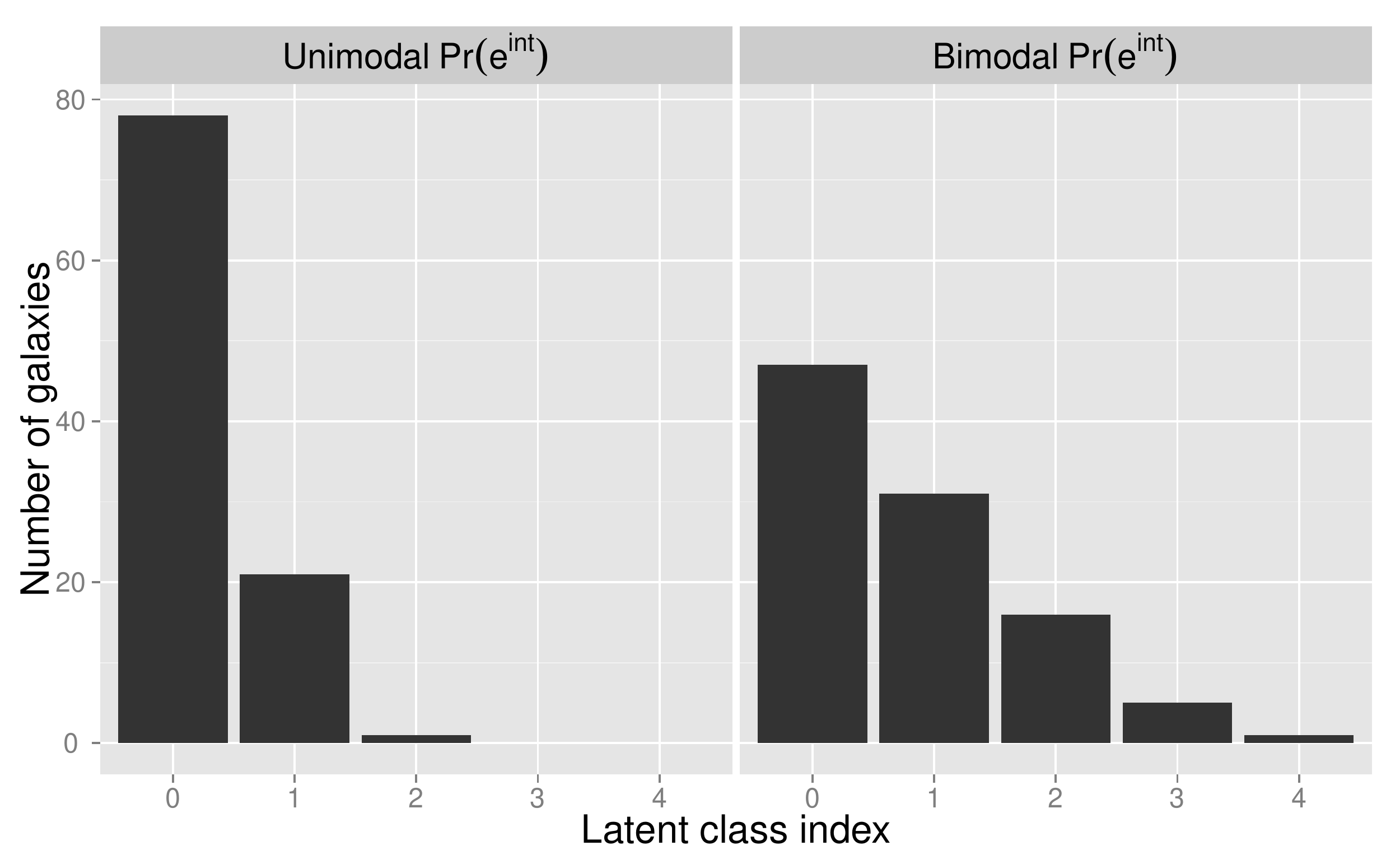}
  }
  \caption{The number of galaxies assigned to each of the DP latent classes for a single sample from  
  the MCMC chains for the `data1' and `data2' mock data scenarios in toy model 2. 
  More galaxies assigned to latent classes with larger indices (starting from 0) indicate 
  more distinct clusters of galaxy ellipticities are favored by the DP model.}
  \label{fig:latent_class_assignments}
\end{figure}
We can also draw some insight about the galaxy population from the DP sampling in the DP model. 
The two peaks in the $\sigma_e$ marginal distribution in \autoref{fig:numerical_example_results} 
indicate the presence of two 
populations of galaxies with different intrinsic ellipticity distributions in the bimodal data set. 
We can draw a similar conclusion from the samples of the latent class selectors in the DP 
model as defined in \autoref{eq:dp2}. \autoref{fig:latent_class_assignments} shows 
the distributions of latent class indices (given integer labels increasing from 0) for a single 
sample from the the DP model MCMC chains.
We see that the posterior samples given the unimodal data set indeed 
favor a single generating distribution while those for the bimodal data set 
favor more than one generating distribution.
The shear inference in all cases is marginalized over these latent class assignments.


\section{Discussion and conclusions} 
\label{sec:conclusions}

\subsection{Hierarchical inference for PSFs and shear} 
\label{sec:discussion}

With the toy models in \autoref{sec:part2} we demonstrated 
how to simultaneously infer a constant shear and the intrinsic galaxy 
ellipticities. We generated interim posterior samples for each 
galaxy independently and then combined the inferences from each galaxy using 
importance sampling and a DP model for the intrinsic ellipticity distribution.
By specifying a common hierarchical conditional PDF on the parameters of the 
intrinsic ellipticity distribution for the galaxy population, we statistically 
correlated, and thereby improved, the marginal shear inference given all galaxies.
A key feature of our model for the intrinsic ellipticity PDF is the assignment 
of galaxies into latent classes based on the inferred intrinsic ellipticity variance.
This allows an improved sensitivity to the action of shear on galaxies with different 
intrinsic ellipticity magnitudes.

Extending the statistical model to that presented in \autoref{eq:full_model} 
to include inference of the PSF and shear that is allowed to vary over the sky 
will have similar effects in correlating the inferences from each galaxy at a 
hierarchical level. 
As an aside, the hierarchical model that imposes parameters common to all 
galaxies in the PDF on, e.g., the intrinsic ellipticity is what allows us 
to perform interim posterior sampling on individual galaxies as in \autoref{sub:importance_sampling}.
Similar conclusions hold for both the 
marginalization of the intrinsic galaxy properties $\galprops$ and the 
lensing potential $\lenspot$.

The shear is correlated across sky positions because of the common gravitational potentials
that distort the light from different galaxies.
Ideally, then, we would like to infer the 3D matter potential $\gravpot$ from the 
combination of galaxy shears and redshift information and infer cosmology from 
the statistics of the 3D mass distribution. While the initial conditions for the 3D 
mass distributions can be modeled as Gaussian distributed to good approximation,
the late-time mass distribution is non-Gaussian and includes significant 
correlations between Fourier modes of the density field. It is not 
obvious how to specify a conditional PDF on the 3D mass distribution
or how to feasibly sample model 
parameters~\citep[see, e.g.,][for an approach to using cosmological $N$-body simulations for this purpose]{schneider2011}. 
\citet{2013MNRAS.432..894J} and \citet{2014arXiv1409.6308J} have presented 
reconstructions of the 3D mass density and cosmological initial conditions from 
galaxy clustering inferred in a spectroscopic survey that might become useful for 
shear inference as well in the future.

If we pursue a description of the mass density in terms of the realization of the 
cosmological initial conditions as in \cite{2014arXiv1409.6308J}, then we would 
infer cosmological parameter constraints by marginalizing the initial conditions $\gravpotinit$.
The lens potential realizations for different source distributions $s$ all depend on the 
same initial conditions realization $\gravpotinit$ evolved under the same cosmology $\thetav$.
So, when we marginalize the initial conditions realization $\gravpotinit$ the distributions 
of $\lenspot_{s}$ are no longer separable for different $s$. This is consistent 
with the common intuition that the lens potentials for, e.g., different photo-$z$ bins, 
should be correlated by common lensing mass along the lines of sight. 
This is also why the probability distribution of $\{\lenspot_s\}$ for all source distributions $s$ 
does not factor over $s$ in \autoref{eq:full_model}. 

\subsection{No more point estimators for galaxy catalogs} 
\label{sub:no_more_postage_stamps}
In traditional algorithms, point estimators for the ellipticity of an individual galaxy 
can be computed from a `postage stamp' image of the galaxy isolated from the 
rest of the survey imaging. The ellipticity estimates from individual galaxies can then 
be averaged over local regions of the sky to estimate the shear. 

Our likelihood in \autoref{eq:likelihood} also applies 
to galaxy `postage stamps'. 
But, there is no statistically meaningful method to combine the information 
from individual postage stamps unless we draw interim posterior samples from 
each postage stamp and use methods such as in \autoref{sub:importance_sampling}. 
We come to an important conclusion then that we should dispense with computing 
point estimators from postage stamp images of galaxies for the purposes 
of later model inferences~\citep[this point was also made previously in][]{2013AJ....146....7B,HoggResearch2012}.



\subsection{Conclusions}
We have described a hierarchical statistical framework to infer cosmic shear from galaxy imaging 
surveys. By explicit specification and marginalization of the statistical models for 
intrinsic (i.e., unlensed) galaxy properties and the PSF, we may be able to improve both the 
accuracy and precision of the shear inference over previously published algorithms.
With a simple toy model in \autoref{sec:part2} we showed promising 
improvements in the precision of the marginal shear inference when galaxy ellipticities 
are generated from a bimodal distribution (intended to mimic the different morphological 
classes of observed galaxies).

We assume a likelihood function for the imaging pixel data. Because we estimate ellipticities and 
shears as galaxy model parameters rather than reduced statistics of the data, defining a 
likelihood for pixel values poses no algorithmic challenges for our framework over using 
reduced statistics.
With an importance sampling algorithm we can infer model parameter constraints
for independent galaxy images and combine these inferences into a consistent shear inference over many 
galaxies. We are only able to perform this importance sampling mathematically by means of a 
hierarchical model that relates the inferences from individual galaxies by means of common 
parameters in the \emph{distributions} over galaxy properties and PSFs. Our framework also 
strongly motivates the use of random samples of posterior model parameters rather than point estimators 
for constructing source catalogs for cosmic shear analysis.

We also introduced a random process to model the distribution of intrinsic galaxy properties 
(such as ellipticity and size). Our Dirichlet Process (DP) model can fit a wider range of 
distributions than any asserted specific functional form.  In \autoref{sub:dp_inference} we
showed that biased inference of, e.g., the intrinsic galaxy ellipticity distribution can be another 
form of `model fitting bias' in the marginal shear constraints. Our DP model helps in alleviating 
the bias in the posterior median shear (e.g., \autoref{tab:shear_bias_ratios}) when the generating 
distribution is multi-modal, but is even more important in reducing the posterior uncertainty on the 
shear (\autoref{fig:numerical_example_results}). 
When we assume a single Gaussian for the generating 
distribution of galaxy ellipticities the DP model yields shear inferences that 
are more biased and no more 
precise than assuming a Gaussian prior on the ellipticities (the Gaussian model with the unimodal 
data set in toy model 2 of \autoref{sub:dp_inference}). 
However, this scenario is a limiting case in that we do not expect 
the distribution of properties of the galaxy population to be well-described by a unimodal Gaussian distribution,
and we do not expect to be able to perfectly match our model distribution to the true generating distribution.
The similar marginal shear inferences we obtain with flat or DP ellipticity variance PDFs 
for the unimodal data set in our toy model 
is therefore a validation of our DP model distribution, which can still perform well when the 
generating distribution for galaxy ellipticities is more complex. An important topic to explore further 
is algorithmic determination of all hyper-prior parameters in the DP model.

While we believe the importance sampling and DP algorithms we presented will be 
useful for the analysis of upcoming large surveys, considerable work remains in implementing variations 
of these algorithms for large-scale imaging analysis. 

\citet{mbi-paper1} shows that our interim posterior 
sampling given galaxy pixel data is computationally limited both by galaxy model parameter optimizations 
and by MCMC sampling. Future developments in fast analytic models for galaxy profiles could reduce the 
computational requirements at this step. The model in~\citet{spergel10} is an intriguing example of such 
model variations worth exploring.
We also speculate that conventional model-fitting biases, resulting from galaxy models that cannot 
describe all important features in the data, may be partially alleviated by a flexible model for the 
PDF on the galaxy model parameters. Because both a change of galaxy model and a change of conditional PDF on a given 
galaxy model are changes in the integration measure when marginalizing intrinsic galaxy properties it 
is possible that a suitable conditional PDF 
specification can reduce remaining model fitting biases.

Our implemention of DP parameter inference relies on Gibbs sampling, which must proceed sequentially in 
all model parameters. This sampling can become a computational bottleneck as the number of latent classes 
increases in the DP (with a corresponding increase in the number of parameters to Gibbs update). 
Parallelizing the Gibbs updates may become useful in cases where the number of latent classes is large. 
However, optimizing the computational performance of the DP sampling is an area for future work.

Our framework also offers interesting possibilities to infer the distribution of intrinsic galaxy 
properties (conditioned on some galaxy modeling assumptions), after marginalizing over the shear and PSF.
Again, work remains to remove any ad-hoc assumptions in the DP model to obtain confidence in any 
inferences about galaxy model parameters that are independent of implicit prior assumptions. That is, we want
want to only impose explicit prior information about the galaxy properties without biases imposed by 
modeling limitations.

Finally, we must still demonstrate the PSF inference and marginalization using our framework with 
pixel-level image simulations. This capability is already part of \tractor code that we use 
in~\citet{mbi-paper1}, but we have not propagated PSF uncertainties to the shear inference.
We are also pursuing methods to incoporate ancillary information about the PSF (e.g., from site-monitoring 
data, wavefront sensors, or engineering images) using the framework in this paper. 


\section*{Acknowledgments}
We thank Dominique Boutigny for several technical reviews of this work and for 
contributions to our GREAT3 challenge submissions based on these methods.
We thank Bob Armstrong, Gary Bernstein, Jim Bosch, and Erin Sheldon for helpful conversations.
We also thank the \greatthree gravitational lensing community challenge team for motivating much of this 
work and providing valuable feedback on the implementation of our algorithms.
Part of this work performed under the auspices of the U.S. Department of Energy by Lawrence Livermore National Laboratory under Contract DE-AC52-07NA27344.
DWH was partially supported by the NSF (grant IIS-1124794) and the Moore-Sloan Data Science Environment at NYU.

\bibliographystyle{apj}
\bibliography{mbi-references}

\appendix

\section{Modeling galaxy images} 
\label{sub:modeling_galaxy_images_the_tractor}

To evaluate the likelihood of the pixel values in \autoref{eq:likelihood} 
we require a computationally efficient code to predict pixel values 
given a galaxy model that can be sheared, convolved with a given PSF, 
and depositied on pixels with the addition of noise. 
\tractor public code\footnote{\url{http://thetractor.org/}} performs exactly 
these operations. 
A key feature of \tractor is that both galaxy flux and PSFs are represented as 
sums of 2D Gaussian distributions allowing fast analytic convolutions of 
model galaxy images with a model PSF. 

For an isolated galaxy image (or ``postage stamp''), we generate posterior samples 
of the galaxy model parameters $\galprops$ via 
MCMC using \tractor to evaluate the likelihood at each MCMC step~\citep{mbi-paper1}.
In all tests of our algorithms we have used the 
public MCMC code\footnote{\url{http://dan.iel.fm/emcee/current/}} 
\emcee~\citep{emcee} for inferring galaxy model parameters and PSFs.

As we discuss in~\citet{mbi-paper1}, the initialization of the galaxy 
model parameters via numerical optimization can be the least accurate 
part of the calculation to infer posterior constraints on a galaxy model.
We have found that an informed choice of `interim prior' on the galaxy 
model can significantly improve the inferences for individual galaxies. 

In the next sub-section we describe how to properly incorporate the posterior samples 
for individual galaxy models with per-galaxy interim 
prior information into the global statistical 
inference for a survey of galaxies. We reserve all further discussion of 
fitting galaxy models to pixel data to \citet{mbi-paper1}.

\section{Choosing a base distribution for the intrinsic galaxy properties} 
\label{sub:base_dist_shear_var}

We can further understand some of the choices in specifying a DP base distribution $G_0$ 
by considering the model where
\begin{equation}
  \pr(\galprops | \alpha) = \normdist_{\galprops}(0, \Sigmamat)  
\end{equation}
such that 
\begin{equation}
  \alpha\equiv \Sigmamat
\end{equation} is the
(co)variance of the distribution of intrinsic galaxy properties.

In the case that $\galprops_n$ is a single parameter so that $\Sigmamat\equiv \sigma^2$,
an obvious conjugate prior for $\sigma^2$ is the inverse-Gamma (IG) distribution. 
However, it is difficult to specify a non-informative prior on $\sigma^2$ using 
the IG distribution when the values of $\sigma^2$ are likely to be much less than one. 
Another choice is the uniform distribution, but this too 
often leads to biased inference of the variance parameter~\citep{gelman06}. 

Instead we follow~\citet{gelman06} and impose a more flexible, but still conditionally conjugate, distribution
on $\sigma^2_{i}$ by decomposing
$\galprops_{i} = \left|\xi_{i}\right| \eta_{i}$, where $\xi$ is a scaling parameter and $\eta_i\sim \normdist(0, \sigma_{\eta}^2)$.
We specify distributions $\xi_{i} \sim \normdist(m, \tau^{-1})$ and $\sigma_{\eta}^{2}\sim \dirproc(\dpprec, G_0)$. 
The standard deviation parameter is $\sigma = |\xi|\sigma_{\eta}$.
The DP base distribution (now for $\sigma_{\eta}^2$) is,
\begin{equation}\label{eq:dp_base_eta_prior}
  G_0(\sigma^2_{\eta}) = \Gamma^{-1}(a_{\eta}, 0.5).
\end{equation}
The family of distributions parameterized by $a_{\eta}, m, \tau$ is 
quite flexible~\citep{Murugiah20121947}. To recover a simpler half-Cauchy PDF we 
take $m=0$ so that
\begin{equation}\label{eq:halfcauchy_prior}
  \pr \left(\sigma^2 | \tau, a_{\eta}\right)  \propto
  \left(1 + \tau \sigma^2\right)^{-(1+a_{\eta})},
\end{equation}
which illustrates how the scaling parameter precision $\tau$ affects the width of the 
base distribution for $\sigma$.

It is possible to generalize the above `variable expansion' to
 a covariance $\Sigmamat$ of a multivariate parameter vector $\galprops_n$ of 
galaxy intrinsic properties. 

\paragraph{Scaling parameter}
To complete the specification of the statistical framework when the 
intrinsic galaxy properties are assumed Gaussian distributed, 
we derive the conditional posterior for $\xi$ after marginalizing over the galaxy 
properties $\galprops$,
\begin{equation}
  \pr(\xi_{n} | \data, \sigma_{\eta, n}, m, \tau, \lenspot) \propto
  \pr\left(\xi_{n} | m,\tau\right)
  \int d\galprops_{n}\, 
  \pr\left(\data_{n}|\galprops_{n} =(\eta_{n}\left|\xi_{n}\right|), \lenspot\right) 
  \pr(\eta_{n} | \sigma_{\eta,n}^2),
\end{equation}
where again we assume $\galprops_n$ is univariate for simplicity of presentation.
Changing integration variables to $\eta_{n}=\galprops_{n}/\left|\xi_{n}\right|$, allows us to use the 
previous importance sampling result~(equation~\ref{eq:importance_samples}) 
for the marginal likelihood after 
including the Jacobian for the variable transformation,
\begin{equation}\label{eq:scaling_param_posterior}
  \pr(\xi_{n} | \data, \sigma_{\eta, n}, m, \tau, \lenspot) \propto
  \pr\left(\xi_{n} | m, \tau\right)
  \frac{1}{N}
  \sum_{k=1}^{N} 
  |\xi_{n}|
  \frac{\prf{\eta_{nk}\equiv \galprops_{nk} / \xi_{n} | \sigma_{\eta,n}^{2}}}
  {\prf{\galprops_{nk} | I_{0}}}.
\end{equation}
We use \autoref{eq:scaling_param_posterior} to conditionally update 
the values of the scaling parameters $\xi_n$ in our MCMC algorithm.

\section{Sampling the DP clustering parameter}
\label{sec:sampling_dp_clustering_parameter}
In many applications (including ours) there is no obvious choice for the
value of the DP clustering parameter $\dpprec$. Instead, we marginalize over 
$\dpprec$ with a suitably chosen prior. 

Following \citet{dorazio09} and \citet{Murugiah20121947} we place a 
conjugate Gamma prior on $\dpprec$ so we can perform Gibbs updates.
\citet{dorazio09} presents a useful algorithm to determine the 
hyperparameters $(a,b)$ of the Gamma prior on $\dpprec$ when little 
information is known about the expected number of clusters in the 
data (i.e., the number of morphologically distinct galaxy populations).

However, given the extensive existing research into galaxy formation 
(e.g., the common classifications of `early' and `late' type galaxies), we
are motivated to place a more informative prior on $\dpprec$.
\citet{Murugiah20121947} describe several algorithms to determine 
the parameters $(a,b)$ for the Gamma prior on $\dpprec$ in informative 
scenarios. 
We can impose our prior knowledge that there are 
early- and late-type galaxies by requiring the expected mean value 
of the number of clusters in the DP to be 2. 
Using the formula from \citet{antoniak1974},
\begin{equation}\label{eq:dp_mean_num_clusters}
  {\rm E}(\numclasses | \ngal, \dpprec) \approx 
  \dpprec \ln \left(\frac{\ngal + \dpprec}{\dpprec}\right),
\end{equation}
for $\ngal \gg 1$, and where $\numclasses$ is the number of clusters or latent classes in the DP.
By solving \autoref{eq:dp_mean_num_clusters} for the value $\bar{\dpprec}$ that gives 
the expected number of clusters, we set the Gamma prior parameters as,
\begin{equation}
  a = \bar{\dpprec} X, \qquad b = X,
\end{equation}
so that the prior mean of $\dpprec$ is $a/b=\bar{\dpprec}$ and the 
prior variance is $a/b^2 = \bar{\dpprec}{X}$ such that larger $X$ 
communicates more certainty on the expected number of morphological classes 
in the galaxy population.
\citet{Murugiah20121947} present other algorithms to encode prior information 
on $\dpprec$ that we reserve for future exploration.

Given the hyperprior parameters $(a,b)$, we perform Gibbs updates of $\dpprec$ with 
the conditional distribution from~\citet{escobar95} (their eq. 13),
\begin{equation}
  \dpprec | \zeta, k \sim
  \pi_{\zeta} \Gamma(a+k, b- \log(\zeta)) 
  + 
  \left(1 - \pi_{\eta}\right)
  \Gamma\left(a + k -1, b - \log(\zeta)\right)
\end{equation}
where $k$ is the number of latent classes currently assigned to the data,
and
\begin{equation}
  \pi_{\zeta} \equiv \frac{a + k - 1}{a + k - 1 + \ngal  (b - \log(\zeta))},
\end{equation}
with
\begin{equation}
  \zeta | \dpprec \sim \beta\left(\dpprec + 1, \ngal\right),
\end{equation}
a variable used only in the Gibbs updates of $\dpprec$, and 
$\beta$ denotes the Beta distribution.

\section{Generating fake posterior samples of galaxy ellipticities}
\label{sec:fake_reaper}
For our `toy model 2' in \autoref{sub:dp_inference} 
we generate fake summary statistics of galaxy pixel data with the following procedure:
\begin{enumerate}
  \item Specify a value for the true reduced shear $g$ for all galaxies (we assume a constant 
  shear for all galaxies in our numerical examples). 
  \item For each galaxy $i=1,\dots,n_{\rm gal}$, draw intrinsic 
  ellipticities from a specified distribution $\shapeint_{i} \sim \pr_{e, {\rm true}}(\cdot)$.
  \item Define the complex ellipticity of the generative model for the pixel data for each 
  galaxy by transforming the intrinsic ellipticity $\shapeint$ with the shear $g$,
  \begin{equation}
    \elensed_{i} = \frac{\shapeint_{i} + g}{1 + g^{*} \shapeint_{i}}.
  \end{equation}
\end{enumerate}

Given the observed ellipticity as a summary statistic of the pixel data, 
we generate interim posterior samples of the galaxy ellipticity by:
\begin{enumerate}
  \item Define an interim prior for the model ellipticity $\pr_{\rm interim}(\elensed_{i})$.
  This prior may be chosen for computational convenience rather than an accurate description of the data, 
  with the knowledge that we will re-weight the posterior samples later to accomodate a hierarchical PDF
  on the intrinsic ellipticity $\shapeint_{i}$.
  \item Define a summary statistic of the pixel data for a galaxy $\summarystat_{i}$ 
  and a functional form for the likelihood $\prf{\summarystat_{i} | \elensed_{i}}$ 
  of the summary statistic.
  \item Draw samples from the interim posterior,
  \begin{equation}
    \pr(\elensed_{i} | \summarystat_{i}) 
    \propto \pr(\summarystat_{i}|\elensed_{i}) \pr_{\rm interim}(\elensed_{i}).
  \end{equation}
\end{enumerate}
The samples of $\elensed_{i}$ contain all the statistical information about galaxy $i$.
For extreme data compression we might limit ourselves to one sample of $\elensed_{i}$ for each 
galaxy.

\section{DP sampling algorithm for toy model 2} 
\label{sec:dp_sampling_algorithm_for_toy_model_2}
\newcommand{\dppost}{\mathcal{H}}

After marginalizing over the intrinsic ellipticity components, the posterior $\dppost_n$ for the intrinsic ellipticity variance given a single galaxy $n$ is (assuming weak shear, $g\ll 1$),
\begin{align}
  \dppost_{n}(\sigma_{\eta}^2 | \xi_{n}, \data_n, g) &=
  G_0(\sigma_{\eta}^2) \int d^2 \eta\, \pr(\data_n | g, \xi, \eta) \cdot \pr(\eta |\sigma_{\eta}^2)
  \notag\\
  &= \frac{(0.5)^{a_{\eta}}}{\Gamma(a_{\eta}} 
  \left(\sigma_{\eta}^2\right)^{-(1+a_{\eta}} \exp \left(-\frac{0.5}{\sigma_{\eta}^2}\right)
  \notag\\
  &\times
  \frac{1}{2\pi \left(\sigma_{\rm obs}^2 + \xi_{n}^{2}\sigma_{\eta}^2\right)}
  \notag\\
  &\times
  \exp \left(-\frac{1}{2}\frac{\left(d_{n,1} - g_{1}\right)^2 + 
    \left(d_{n,2} - g_{2}\right)^2}{\sigma_{\rm obs}^2 + \xi_{n}^{2}\sigma_{\eta}^2}\right).
\end{align} 
This expression is pedagogical only. 
In practice, we do not have $\data_n$ values.
Instead we have importance samples of $\shapeint \equiv |\xi|\eta$ with which we perform the 
above integral via Monte Carlo.

Following from \autoref{eq:dp_cond_prior} and \citet{neal00}, 
the conditional posterior for $\sigma^2_{\eta,n}$ for galaxy $n$ is,
\begin{equation}
  \sigma^2_{\eta,n} | \sigma^2_{\eta,-n}, \data_n 
  \sim 
  \sum_{m \neq n} 
  q_{n,m} \delta_{\rm D}\left(\sigma^2_{\eta,m}\right) + 
  r_n \dppost_n
\end{equation}
where $\delta_{\rm D}(\sigma^2_{\eta,m})$ is a Dirac delta function indicating that $\sigma_{\eta,n}^2$ for galaxy $n$ takes the same value as that for galaxy $m$. This is how clustering occurs in the Gibbs sampling. The coefficient of the delta functions $q_{n,m}$ is defined to be the marginal likelihood for galaxy $n$ given the parameter $\sigma^2_{\eta,m}$ currently assigned to galaxy $m$,
\begin{equation}
  q_{n,m} = b \int d^2\eta\, \pr(\data_n | \eta, \xi_{m}, g) \cdot \pr(\eta| \sigma^2_{\eta,m}),
\end{equation}
where $b$ is a constant defined below. 
If all the $q_{n,m} \forall m \neq n$ are sufficiently small, 
meaning that no parameters $\sigma^2$ currently assigned to other galaxies are good descriptions of 
observation $n$, then we draw a new value for $\sigma_{\eta,n}^2$ from $H_n$. 
The coefficient that defines the probability for drawing new values relative to the 
coefficients $q_{n,m}$ is the likelihood for galaxy $n$ marginalized over both the intrinsic ellipticity components and the hyperprior parameter $\sigma_{\eta}^2$,
\begin{equation}
  r_n = b\, \dpprec \int d\sigma_{\eta}^2\, \int d^2\eta\, 
  \pr(\data_n | \eta,\xi_{n}, g) \cdot 
  \pr(\eta| \sigma_{\eta}^2) \cdot 
  G_0(\sigma_{\eta}^2 | a_{\eta}).
\end{equation}
Note that $r_n$ is proportional to $\dpprec$ so that larger $\dpprec$ values mean 
less clustering of $\sigma_{\eta,n}^2$ values for each galaxy.

The parameter $b$ is defined so that 
\begin{equation}
  \sum_{m\neq n}q_{n,m} + r_n = 1.
\end{equation}

Given the importance samples for $\shapeint$, we compute $q_{n,m}$ as,
\begin{equation}
  \frac{1}{b} q_{n,m} = \frac{1}{K} \sum_{k=1}^{K} 
  \frac{\pr(\shapeint_{kn}/\xi_{n} | \sigma^2_{\eta,m})} {\pr_{\rm interim}(\shapeint_{kn})}
\end{equation}

Similarly, we compute $r_n$ by first integrating analytically over $\sigma_{\eta}^2$ and then summing over the $\shapeint$ samples, 
\begin{equation}
  \frac{1}{b}r_{i} = \frac{1}{K} \sum_{k=1}^{K}
  \frac{\pr_{\rm marg}(\shapeint_{kn} / \xi_{n} | a_{\eta})}{\pr_{\rm interim}(\shapeint_{kn})},
\end{equation}
where, assuming the parameter $m=0$ as in \autoref{eq:halfcauchy_prior},
\begin{align}
  \pr_{\rm marg}\left(\eta | a_{\eta}\right) 
  &\equiv \frac{0.5^{a_{\eta}}}{2\pi} 
  \frac{\Gamma(1 + a_{\eta})}{\Gamma(a_{\eta})}
  \left(\frac{1}{2}\left(1 + |\eta|^2\right)\right)^{-(1 + a_{\eta})}
\end{align}
and $a_{\eta}$ is a parameter in the inverse Gamma base distribution $G_0$ as 
defined in \autoref{eq:dp_base_eta_prior}.


\section{Galaxy moments as summary statistics of the pixel data} 
\label{sec:galaxy_moments}
\newcommand{\shearmean}{\bar{\gamma}}
\newcommand{\gpcov}{\mathsf{\Sigma}}
\newcommand{\qv}{\mathbf{Q}}
\newcommand{\qvest}{\hat{\qv}^{\rm ML}}
\newcommand{\lpf}{\tilde{\bm \lenspot}}
\newcommand{\linmap}{\mathsf{G}}
\newcommand{\shearvec}{{\bm \Gamma}}
\newcommand{\npix}{n_{\rm pix}}
\newcommand{\onesmat}{\mathsf{1}}
\newcommand{\sigmaqp}{\Sigma_{Q,{\rm prior}}}
\newcommand{\ilim}{I_{\rm lim}}
\newcommand{\xlim}{x_{\rm lim}}
\newcommand{\qvobs}{\mathbf{Q}^{\rm obs}}
\newcommand{\qvobsh}{\hat{\mathbf{Q}}^{\rm obs}}
\newcommand{\qvint}{\mathbf{Q}^{\rm int}}
\newcommand{\ntemp}{N_{\rm temp-gal}}
\newcommand{\re}{r_{e,\nu}}



\citet{Bernstein+Armstrong2013} (BA14) recently enumerated the challenges in modeling 
galaxy images at the pixel level
because of the sensitivity to unknown features in complex galaxy morphologies. 
Throughout this paper we have assumed we can find a parameterized 
model for the galaxy morphologies that is sufficiently descriptive and 
flexible to avoid large model fitting biases. However we have not 
described and validated such a model and it remains possible that model-fitting 
biases will remain a significant challenge within our statistical framework.

In this appendix we consider the approach in~BA14 
and assume the galaxy pixel data is compressed into a set of moments.
We will demonstrate that there can be stringent requirements on the 
prior knowledge of the galaxy moments that are not obviously easier to 
satisfy than the pixel-level model specification we assume in the main 
body of our paper.

In the simplest analysis we may consider just the quadrupole moments of the intensity distribution,
\begin{equation}\label{eq:second_moments}
  Q_{ij} \equiv \int d^2\fc\, H_i\, H_j\, 
  I(\fc)
  W(|\fc|),
\end{equation}
where $\fc$ are image plane coordinates, $I(\fc)$ is the intensity distribution 
of a galaxy, and
$W(|\fc|)$ is an arbitrary weight function introduced to exclude 
regions of the galaxy image that may be noise-dominated (e.g., at large 
radii with faint surface brightness and on small spatial scales 
where the PSF removes morphological information).

Next we summarize how the galaxy moments change under lensing.
The two shear components and the magnification that can be estimated from a galaxy image 
are derived from components of the Hessian of the lensing potential $\lenspot$,
\begin{align}
  \mathsf{A}(\fc) &\equiv \left[\delta_{ij} - 
  \frac{\partial^2\lenspot(\fc)}{\partial H_{i}\partial H_{j}}\right]
  = 
  \left(
  \begin{array}{cc}
  1 - \kappa - \shear_1 & -\shear_2 \\
  -\shear_2 & 1 - \kappa + \shear_1
  \end{array}
  \right)
  \label{eq:lens_matrix_def}\\
  &= (1 - \kappa)
  \left(
  \begin{array}{cc}
  1 - g_1 & -g_2 \\
  -g_2 & 1 + g_1
  \end{array}
  \right) 
\end{align}
where $g_{i}\equiv \shear_{i}/(1-\kappa)$ is the reduced shear, 
which is approximately equal to the 
shear when $\kappa \ll 1$ as in the weak lensing regime.

The intensity distribution of a lensed image can be mapped to that of the 
unlensed image via the distortion matrix $\mathsf{A}$ in a 
linear model in the weak lensing regime~\citep[][eq. 3.13]{bartelmann01},
\begin{equation}\label{eq:lens_effect_on_intensity}
  I(\fc) = I^{(s)}\left[\fc_{0} + \mathsf{A}(\fc_{0} 
  \left(\fc - \fc_{0}\right) \right].
\end{equation}
Using equation~\ref{eq:lens_effect_on_intensity} the 
lensed moments can be related to the unlensed moments via,
\begin{align}\label{eq:lensed_moments}
  \mathsf{Q}^{\rm obs}_{ij} &= \int d^{2} \fc\, 
  H_{i} H_{j} 
  \left[I(\mathsf{A}\fc) \ast P_{\rm pix}(\fc)\right]
  W(|\fc|)
  \notag\\
  &= \left|\mathsf{A}\right|^{-1} \int d^2 H'\,
  \left(\mathsf{A}^{-1}\fc'\right)_{i} \left(\mathsf{A}^{-1}\fc'\right)_{j}
  \left[I(\fc') \ast P_{\rm pix}\left(\mathsf{A}^{-1} \fc'\right)\right]
  W\left(\left|\mathsf{A}^{-1}\fc'\right|\right)
  \notag\\
  &\approx \left|\mathsf{A}\right|^{-1} 
  \mathsf{A}^{-1} \mathsf{Q}^{\rm int} \mathsf{A}^{-1}
\end{align}
where, without loss of generality, we have assumed $\fc_{0}=\mathbf{0}$, 
we introduced the pixel window function $P_{\rm pix}$ that is convolved 
with the galaxy intensity profile, the symbol $\ast$ denotes convolution, 
and the final equality holds under the assumptions of weak shear 
(such that $\left|\mathsf{A}^{-1}\fc\right|\approx|\fc|$) and
well-sampled galaxies (so that changes in the pixel shape have little 
impact on the observed intensity profile).
Under these assumptions, \autoref{eq:lensed_moments} gives a linear 
relation between the lensed and unlensed moments.

We will assume this linear relation between the lensed and unlensed moments 
in order to analytically marginalize the intrinsic galaxy moment parameters below.
Dropping the assumption of small pixels should only 
increase the requirements on the knowledge of the unlensed galaxy moments while 
non-weak shear has been a negligible correction in previous 
cosmic shear analyses~\citep[e.g.][]{jee2013}.
The linear relation in the final equality of 
\autoref{eq:lensed_moments} will also be broken by surface brightness 
or SNR cuts as well as deconvolution of an anisotropic PSF, but 
we also neglect these issues in the subsequent analysis.

BA14 and \citep{Bernstein2010} advocate the use of 
galaxy moments defined in Fourier space; what they call `Bayesian Fourier 
Domain' (BFD) moments,
\begin{equation}\label{eq:bfd_moments}
  M_{r, +, \times} \equiv
  \int d^2 k\, 
  \left(
  \begin{array}{c}
  k_x^2 + k_y^2\\
  k_x^2 - k_y^2\\
  2 k_xk_y
  \end{array}
  \right)
  \left[I(\kv) \ast W_{\rm apod}(\kv)\right]
  W(k),
\end{equation}
where $I(\kv)$ is assumed to be the observed intensity after deconvolving the PSF (which 
is a simple division operation in Fourier space) and $W(k)$ is now 
an isotropic weight 
that must drop to zero for wavenumbers where the Fourier transfrom of the PSF, 
the modulation transfer function (MTF) is small 
(to avoid including excess noise in the estimator for the moments).
The effect of lensing on the BFD moments is,
\begin{equation}\label{eq:BFDtransform}
  M_{ij} = \int d^{2}k\, \left(A \kv \right)_{i} \left(A\kv \right)_{j}
  W \left(\left|A\kv\right|\right) I(\kv).
\end{equation}
So the Fourier-domain moments transform as 
$\mathsf{M}^{\rm obs}\approx\mathsf{A}\mathsf{M}^{\rm int}\mathsf{A}$ while the 
configuration-space moments transform as 
$\mathsf{Q}^{\rm obs}\approx|\mathsf{A}|^{-1}\mathsf{A}^{-1}\mathsf{Q}^{\rm int}\mathsf{A}^{-1}$.

Alternatively, we could arrange the 3 independent moments into 
the data vector,
$ \qv \equiv \left(Q_{11}, Q_{12}, Q_{22}\right)$. Then we can 
rewrite equation~\ref{eq:lensed_moments} as,
\begin{equation}
  \qvobs \equiv \mathsf{S} \qvint
\end{equation}

The linear moments lensing operator 
defined in terms of the convergence and two shear components is,
\begin{equation}\label{eq:lin_op}
  \mathsf{S} \equiv
  \frac{1}{\left[(1-\kappa)^2 - \shear_1^2 - \shear_2^2\right]^{3}}
  \left(
  \begin{array}{ccc}
  1 + 2 g_1 + g_1^2 + g_2^2 & 2g_2 & 0 \\
  2 g_2 & 1 + g_1^2 + g_2^2 & 2 g_2 \\
  0 & 2g_2 & 1 - 2 g_1 + g_1^2 + g_2^2
  \end{array}
  \right)
\end{equation}
For a small shear,
\begin{equation}\label{eq:lin_op_weak_shaer}
  \mathsf{S} \approx
  \left(
  \begin{array}{ccc}
  1 + 2 \shear_1 & 2 \shear_2 & 0 \\
  2 \shear_2 & 1 & 2 \shear_2 \\
  0 & 2 \shear_2 & 1 - 2 \shear_1
  \end{array}
  \right) 
\end{equation}
The analogous linear operator for the BFD moments is different, but 
we will abuse our notation to use $\mathsf{S}$ to indicate either operator 
below. For the BFD moments,
\begin{equation}\label{eq:lin_op_bfd}
  \mathsf{S} \equiv
  (1-\kappa)^2
  \left(
  \begin{array}{ccc}
  1 - 2 g_1 + g_1^2 + g_2^2 & 2g_2 & 0 \\
  2 g_2 & 1 + g_1^2 + g_2^2 & 2 g_2 \\
  0 & 2g_2 & 1 + 2 g_1 + g_1^2 + g_2^2
  \end{array}
  \right).
\end{equation}

\subsection{Statistical model} 
\label{sub:statistical_model}
We assume a Gaussian likelihood for the observed second moments, which 
we expect to be a reasonable approximation when the noise is dominated by weakly correlated 
pixel noise~(BA14, and \autoref{sub:likelihood_specification}), 
\begin{align}\label{eq:moments_likelihood}
  \prf{\qvobsh |\qvint, \mathsf{S}} 
  &= \frac{1}{\left|\Sigma_{Q}\right|^{1/2} (2\pi)^{3/2}}
  \exp \left[-\half 
  \left(\qvobsh - \mathsf{S}\qvint\right)^{T}
  \Sigma_{Q}^{-1}
  \left(\qvobsh - \mathsf{S}\qvint\right)\right],
\end{align}
where the hat denotes the observed values rather than the model prediction 
for the observations. The covariance $\Sigma_{Q}$ allows for propagating 
an arbitrary (correlated) pixel noise model.

\subsubsection{The galaxy shape prior} 
\label{sec:models_for_galaxy_shape_priors}

For marginalizing over the intrinsic galaxy shape moments, it is convenient 
to rewrite the likelihood explicitly as a multivariate Gaussian in $\qvint$,
\begin{equation}
  \prf{\qvobsh | \qvint, \mathsf{S}} 
  \propto
  \frac{1}{\left|\Sigma_{Q}\right|^{1/2}} 
  \exp\left[-\half 
  \left(\qvint - \qvest \right)^{T} 
  \mathsf{S}^{T} \Sigma_{Q}^{-1} \mathsf{S}
  \left(\qvint - \qvest \right)
  \right]
where
\end{equation}
\begin{equation}
  \qvest \equiv \mathsf{S}^{T} \left(\mathsf{S}\mathsf{S}^{T}\right)^{-1} \qvobsh
\end{equation}
is the maximum-likelihood estimator for the intrinsic (i.e., unlensed) 
galaxy shape moments.

\subsubsection{Gaussian prior} 
\label{sub:gaussian_prior}
In the limit of strongly informative prior on the intrinsic galaxy shape, 
we can approximate the prior as a Gaussian distribution centered on the 
assumed galaxy shape moments.

With a conjugate Gaussian prior on $\qvint$ with mean $\muv_{Q}$ and 
covariance $\sigmaqp$, the marginal posterior distribution for the 
lensing quantities $\mathsf{S}$ is,
\begin{align}
  \prf{\mathsf{S}|\qvobsh} 
  &\propto 
  \frac{\left|\mathsf{S}^{T}\Sigma_{Q}^{-1}\mathsf{S} + \sigmaqp^{-1}\right|^{-1/2}}
  {\left|\left(\mathsf{S}^{T}\Sigma_{Q}^{-1}\mathsf{S}\right)^{-1} + \sigmaqp\right|^{1/2}}
  \notag\\
  &\qquad\times
  \exp\left[-\half
  \left(\qvest - \muv_{Q}\right)^{T}
  \left(\left(\mathsf{S}^{T}\Sigma_{Q}^{-1}\mathsf{S}\right)^{-1} + \sigmaqp\right)^{-1}
  \left(\qvest - \muv_{Q}\right)
  \right]
  \notag\\
  &= 
  \left| 2\ident + \left(\mathsf{S}\sigmaqp\mathsf{S}^{T}\right) \Sigma_{Q}^{-1}
  + \Sigma_{Q} \left(\mathsf{S}\sigmaqp\mathsf{S}^{T}\right)^{-1}
  \right|^{-1/2}
  \notag\\
  &\qquad\times
  \exp\left[-\half
  \left(\qvobsh - \mathsf{S}\muv_{Q}\right)^{T} 
  \left(\Sigma_Q + \mathsf{S}\sigmaqp\mathsf{S}^{T}\right)^{-1}
  \left(\qvobsh - \mathsf{S}\muv_{Q}\right)
  \right]\label{eq:shear_posterior}.
\end{align}
This is a nontrivial shear posterior both 
because $\mathsf{S}$ is nonlinear in the shear and $\mathsf{S}$ 
appears in both the normalization and the exponent. 
It is worth recalling that we arrived at this result with a simple 
set of assumptions, namely a Gaussian likelihood and intrinsic shape prior
and a linear relation between the lensed and unlensed galaxy moments.
Our strongest assumption is possibly that of a Gaussian prior on the 
intrinsic moments. But, if we are willing to consider more general 
priors that are composed of sums of Gaussians, or the limit of large variance in the Gaussian prior,
then we might expect to gain valuable intuition and forecasts by 
examining the implications of the shear posterior in 
equation~\ref{eq:shear_posterior}. We follow this line of inquiry in the 
next section.

\paragraph{Gaussian limit for the shear posterior} 
BA14 assumed a Gaussian posterior distribution for the shear. 
Our shear posterior in equation~\ref{eq:shear_posterior} approaches a Gaussian form 
when,
\begin{enumerate}
  \item $\mathsf{S}$ is linear in the shear, as in equation~\ref{eq:lin_op_weak_shaer}.
  \item $z^{T}\Sigma_{Q} z \gg z^{T}\mathsf{S}\sigmaqp\mathsf{S}z \qquad \forall z\in R^{3n}$, where $n$ is the number of observed galaxies.
\end{enumerate}
These conditions can be understood as the (1) weak shear and (2) strong shape prior regimes.
We have already made these assumptions to some degree in assuming 
a linear relation between the lensed and unlensed moments and the
Gaussian form for the prior on the moments.

\subsection{Impact of the shape prior} 
The Gaussian shape prior has two parameters: a mean and covariance. 
A choice of shape prior mean that is different from the truth for a given 
galaxy will introduce a bias in the shear inference with a magnitude that depends on the 
width of the prior. For a given mean, different 
shape prior covariance choices will make the prior more or less informative. 

We use the integral of the marginal shear posterior, also known as the Bayesian evidence,
\begin{equation}\label{eq:evidence}
  \prf{\qvobs, \sigmaqp, \muv_{Q}|\thetav}
  \int d\kappa\, d\shear_1\,d\shear_2\,
  \prf{\mathsf{S}(\kappa,\shear_1,\shear_2) | \qvobs, \sigmaqp, \muv_{Q}} 
  \prf{\kappa,\shear_1,\shear_2 | \thetav}
\end{equation}
to characterize how the choice of shape prior parameters affects the shear inference. 
\autoref{eq:evidence} can also be understood as the probability of the data given the 
shape prior parameters for all possible shear values, for a specified 
shear prior $\pr(\kappa, \shear_1, \shear_2|\thetav)$, with $\thetav$ 
the cosmological parameters.

\begin{figure}
  \centerline{
    \includegraphics[width=0.7\textwidth]{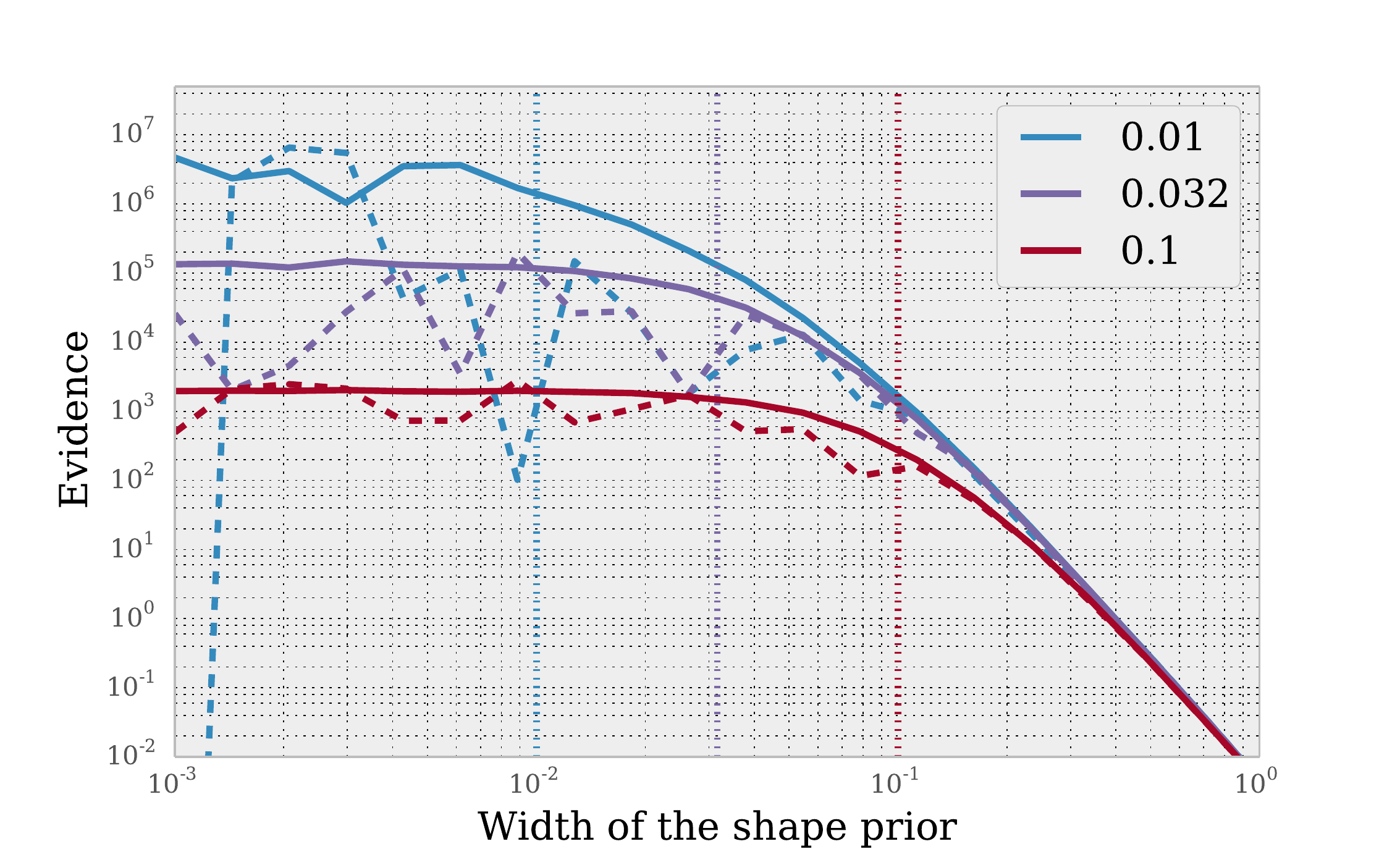}
  }
  \caption{Bayesian evidence of the shear posterior as a function of the assumed 
  width in the galaxy shape prior in units of the galaxy intensity second moments.
  The different colors show different assumed values for the observational error 
  (in units of the observed moments). The vertical lines denote the same 3 observational 
  error values to offer a comparison with the peak locations in the colored lines.
  In all cases, the evidence reaches a maximum for values of the shape prior width 
  (e.g., $\sqrt{\sigmaqp}$) that is within a few times the observational error, with little 
  information gained as the shape prior width decreases below the observational error.
  The solid lines assume no bias in the mean of the shape prior. The dashed lines 
  show the evidence after marginalizing over the bias in the mean of the shape prior, 
  with a Gaussian hyper-prior of width 0.07.
  }
  \label{fig:evidence}
\end{figure}

We plot the evidence as defined in \autoref{eq:evidence} in \autoref{fig:evidence} 
as a function of the `width' $\sigma$ of the shape prior assuming $\sigmaqp$ is diagonal 
with equal variance values $\sigma^2$ on the diagonal. The prior width $\sigma$ is plotted 
in units of the second moments.
Note also that this example assumes only one of many possible 
values for the intrinsic shape moments.
\label{sec:exploration}
The evidence increases with decreasing shape prior width until the prior is of similar width 
to the observational errors in the galaxy moments (denoted by the vertical lines in 
\autoref{fig:evidence}. This is a qualitative 
result that can be understood 
with intuition, but Fig.~\ref{fig:evidence} now quantifies the information lost when the 
shape prior is a specific size relative to the observational errors and the observed moments.

We can also explore the requirements on specification of the mean of the shape prior by 
marginalizing over $\mu_Q$ in \autoref{eq:evidence} with a Gaussian hyperprior on 
$\mu_Q$ with mean equal to the true value (again denoted $\mu_Q$ in a convenient mangling 
of notation). The $\mu_Q$ marginalized results are shown by the dashed lines in \autoref{fig:evidence}.
We assume a shape prior mean hyperprior width of $0.07$ to marginalize over $\mu_Q$ (i.e., 
the mean of the prior on the moments is known to within $7\%$ of the moment values). 
In this case, the evidence assuming an observational 
error of 0.01 (blue dashed line) is similar to that assuming an observational error of 0.03 (purple dashed line). 
So, we might 
tolerate an unknown bias in the shape prior mean of up to 7\% as long as the observational 
errors on the moments are $\sim 3$ times smaller than we would require if the shape prior mean were 
known perfectly.

While this numerical example is contrived, it illustrates the difficulty in constructing a 
scenario in which prior knowledge of the moments is not important.

\end{document}